\documentclass[useAMS,usenatbib]{mnras}

\usepackage{graphicx}	% Including figure files
\usepackage{amsmath}	% Advanced maths commands
\usepackage{amssymb}	% Extra maths symbols
\usepackage{savesym}
\usepackage{mathrsfs,dsfont}
\usepackage{multirow}
\usepackage{captcont}
\usepackage{float}
\usepackage{epstopdf}
\usepackage{color}
\graphicspath{{figures/}}

\def\be{\begin{equation}}
\def\ee{\end{equation}}
\def\kms{{\rm \,km\,s^{-1}}}

\def\Gyr{{\rm \,Gyr}}

\def\kpc{{\rm \,kpc}}

\def\keV{{\rm \,keV}}

\newcommand{\chandra}{{\em Chandra} }
\newcommand{\athena}{{\em ATHENA} }
\newcommand{\lynx}{{\em LYNX} }
\newcommand{\rosat}{{\em ROSAT} }
\newcommand{\BV}{{Brunt-V\"{a}is\"{a}l\"{a}}}
\newcommand{\Fr}{{\rm Fr}}

\title[Buoyant Bubbles in Galaxy Clusters]{Generation of Internal Waves by Buoyant Bubbles in Galaxy Clusters and Heating of Intracluster Medium}

\author[Congyao Zhang et al.]{
Congyao Zhang,$^1$\thanks{E-mail: cyzhang@mpa-garching.mpg.de}
Eugene Churazov,$^{1,2}$
Alexander A. Schekochihin$^{3,4}$
\\
$^1$~Max Planck Institute for Astrophysics, Karl-Schwarzschild-Str. 1, D-85741 Garching, Germany  \\
$^2$~Space Research Institute (IKI), Profsoyuznaya 84/32, Moscow 117997, Russia \\
$^3$~Rudolf Peierls Centre for Theoretical Physics, University of Oxford, Oxford OX1 3NP, UK \\
$^4$~Merton College, Oxford OX1 4JD, UK
}

\date{Accepted XXX. Received YYY; in original form ZZZ}

\pubyear{2018}

\begin{document}
\label{firstpage}
\pagerange{\pageref{firstpage}--\pageref{lastpage}}
\maketitle

\begin{abstract}
Buoyant bubbles of relativistic plasma in cluster cores plausibly play a key role in conveying the energy from a supermassive black hole to the intracluster medium (ICM)~-~the process known as radio-mode AGN feedback. Energy conservation guarantees that a bubble loses most of its energy to the ICM after crossing several pressure scale heights. However, actual processes responsible for transferring the energy to the ICM are still being debated. One attractive possibility is the excitation of internal waves, which are trapped in the cluster's core and eventually dissipate.
Here we show that a sufficient condition for efficient excitation of these waves in stratified cluster atmospheres is flattening of the bubbles in the radial direction. In our numerical simulations, we model the bubbles phenomenologically as rigid bodies buoyantly rising in the stratified cluster atmosphere. We find that the terminal velocities of the flattened bubbles are small enough so that the Froude number $\Fr\lesssim 1$. The effects of stratification make the dominant contribution to the total drag force balancing the buoyancy force. In particular, clear signs of internal waves are seen in the simulations. These waves propagate horizontally and downwards from the rising bubble, spreading their energy over large volumes of the ICM. If our findings are scaled to the conditions of the Perseus cluster, the expected terminal velocity is $\sim100-200\kms$ near the cluster cores, which is in broad agreement with direct measurements by the Hitomi satellite.
\end{abstract}


%%%%%%%%%%%%%%%%%%%%%%%%%%%%%%%%%%%%%%%%%%%%%%%%%%
%%%%%%%%%%%%%%%%% BODY OF PAPER %%%%%%%%%%%%%%%%%%

\section{Introduction} \label{sec:introduction}
Radio mode of active galactic nucleus (AGN) feedback has been widely accepted as the mechanism to prevent the gas in the cores of galaxy clusters from catastrophic overcooling (see, e.g., \citealt[][]{Fabian2012,McNamara2012,Vikhlinin2014,Soker2016} for reviews). Clear signs of the intracluster medium (ICM) interaction with radio lobes inflated by radio jets from supermassive black holes (SMBHs) have already been seen in \rosat images of the Perseus and M87/Virgo clusters \citep[][]{Boehringer1993,Boehringer1995}. The very same images provide an estimate (lower limit) of the power needed to inflate these bubbles based on the comparison of the inflation and buoyancy time scales and show that this power is comparable to the gas cooling losses  \citep{Churazov2000}. This conclusion has been confirmed with much sharper \chandra images \citep[e.g.,][]{McNamara2000,Fabian2000} and extended for large samples of clusters \citep[e.g.,][]{Birzan2004,Hlavacek2012}. Radio emission from the bubbles implies that at least part (and plausibly most) of the bubble pressure support is provided by Cosmic Rays (CRs) and magnetic fields. If CRs are confined within the bubbles, then the interaction with the ICM is purely mechanical, mediated by the motion of bubble boundaries.
An expanding boundary can launch shocks and sound waves into the ICM, if it expands quickly. For a very rapid expansion, most of the energy goes into a shock-heated gas shell, while the fraction of the energy carried away by an outgoing sound wave is less than $\sim$12.5\% if the boundary expansion is spherically symmetric \citep{Tang2017}. In the opposite limit of a relatively slow expansion, most of the energy released inside the bubble is stored in the form of the bubble enthalpy (the sum of internal thermal energy and $PV$ work done by the boundary). Here we concentrate on the latter scenario, in which the bubble enthalpy accounts for most of the SMBH energy output.

Bubbles of relativistic plasma are buoyant in the cluster atmosphere and will move to larger radii \citep{Gull1973}. Energy conservation arguments imply that much of a bubble's enthalpy will be transferred to the ICM once the bubble crosses several pressure scale heights \citep[][]{Churazov2001,Churazov2002,Begelman2001}. While these arguments guarantee high coupling efficiency of the radio-mode AGN feedback, they do not depend on the properties of the ICM and, therefore, by themselves do not single out a particular process responsible for the energy transfer to the ICM. One can therefore pose a question as to the nature of the drag force that balances the buoyancy, once the bubble, which is assumed to be essentially massless, reaches its terminal velocity. For instance, it could be viscous or magnetic stresses, turbulence generated in the wake of the bubbles, advection of low-entropy gas to large radii, excitation of sound waves or internal waves. The latter possibility is attractive on several grounds. First, internal waves are trapped in the central region of a cluster, because the {\BV} frequency ($N$) is a decreasing function of radius \citep[][]{Balbus1990,Lufkin1995}, implying that the energy will not leak outside the cluster core. Secondly, these waves can travel in the tangential direction (azimuthal) and spread the energy throughout the core.  The question then arises: how efficient is the excitation of internal waves? This is the focus of this study.

A comprehensive answer to this question requires detailed knowledge of the ICM micro-physics and the internal composition and structure of the bubbles, which is currently missing. Numerical methods face a serious challenge in modeling bubble dynamics and understanding the relevant heating processes. More specifically, ideal hydrodynamic models lead to rapid destruction of rising bubbles. Boundaries of buoyant bubbles are susceptible to the Rayleigh–-Taylor (R-T), Kelvin–-Helmholz (K-H) and Richtmyer-–Meshkov (R-M) instabilities. Therefore, it is hard for the bubbles to survive for longer than $\sim0.1-0.2\Gyr$ in hydrodynamic simulations, unless high viscosity and/or magnetic fields suppress the interface instabilities \citep[e.g.,][]{Reynolds2005,Dong2009,Bambic2018}. Observations show that, however, some clusters (e.g., Perseus, M87/Virgo) have X-ray cavities with relatively regular shapes even far from the cluster center \citep{Fabian2011,Forman2007}. Phenomenologically, this implies that a strong surface tension acts on the bubble surface and keeps the bubble stable, although the detailed physical description of this effective surface tension, presumably magnetic, is difficult. To circumvent these problems, we make two major simplifying assumptions: $\rm i)$ the ICM can be described in the framework of ideal hydrodynamics and $\rm ii)$ the shape of the bubble does not change as the bubble rises in the cluster atmosphere. Thus, we model the bubbles phenomenologically as rigid bodies buoyantly rising in a stratified cluster atmosphere and discuss the nature of the drag force acting on these bubbles.

This paper is organised as follows. In Section~\ref{sec:method}, we describe the models and simulation methods adopted in this work. The main results from our simulations are presented in Section~\ref{sec:results}, where we study systematically the generation of internal waves by bubbles of different shapes rising in a stratified atmosphere. In Section~\ref{sec:discussions}, we discuss the implications of our findings for the terminal velocity of the bubbles in galaxy clusters, and velocity diagnostic with future high energy resolution missions. In Section~\ref{sec:conclusions}, we summarize our conclusions.

\section{Preliminary considerations} \label{sec:preliminary}

\begin{figure}
\centering
\includegraphics[width=0.45\textwidth]{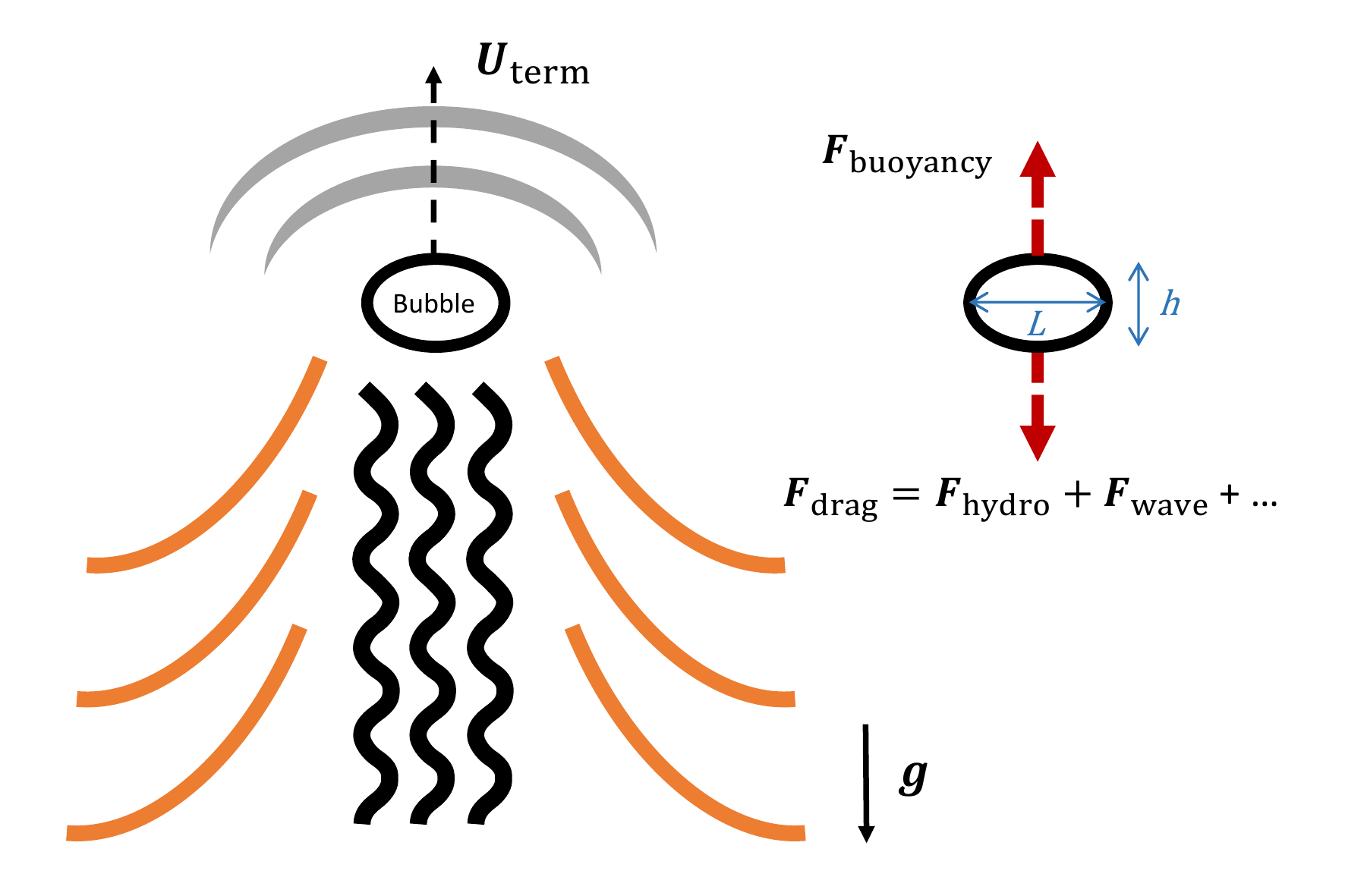}
\caption{Sketch showing a bubble rising in a stratified medium. The bubble rises at the terminal velocity when the buoyancy force is balanced by the drag force. The gray, black and orange lines show schematically sound waves, turbulence, and internal waves excited by the moving bubble, which all can contribute to the total drag.}
\label{fig:sketch}
\end{figure}

In the scenario of a quasi-continuous radio-mode AGN feedback, the generation of X-ray cavities in the centers of galaxy clusters by the SMBH can be generally summarized as a two-stage process. First, a pair of bubbles are blown by bipolar jets, and subsequently expand until the expansion velocity becomes comparable to the velocity of their rise driven by the buoyancy force. At that moment, the expansion is subsonic and the bubble is close to pressure equilibrium with the surrounding ICM. Second, the relic bubbles detach from the cluster center and buoyantly rise upwards. The bubbles finally reach their terminal velocity when drag balances buoyancy force as shown in Figure~\ref{fig:sketch}.

Consider a stably stratified atmosphere in hydrostatic equilibrium, which is locally characterized by the {\BV} frequency
\begin{eqnarray}
N=\sqrt{\frac{g}{\gamma H_{\rm S}}}=\frac{c_s}{\gamma\sqrt{H_{\rm S}H_{\rm P}}},
\label{eq:bv_frequency}
\end{eqnarray}
where $g$, $\gamma$ and $\displaystyle c_s$ are the gravitational acceleration, gas adiabatic index, and adiabatic sound speed, respectively; $H_{\rm S}=|{\rm d}\ln S /{\rm d} r|^{-1}$, where $S$ is the gas entropy; $H_{\rm P}=|{\rm d}\ln P /{\rm d} r|^{-1}$, where $P$ is the gas pressure. Such an atmosphere supports sound and internal waves, the latter with frequencies $\omega$ below $N$, while the former have frequencies above the so-called acoustic cutoff frequency $\omega_{\rm a}=c_s/2H_\rho\gtrsim N$, where $H_{\rm \rho}=H_{\rm P}$ is the density scale height in an isothermal atmosphere.

Now consider a buoyant bubble, whose mass can be neglected, rising with terminal velocity $U_{\rm term}$ in such an atmosphere. Let us first neglect the stratification and the compressibility of the gas and assume that the drag is purely hydrodynamic, i.e.,
\be
F_{\rm buoyancy}(\equiv g\rho V_{\rm b})=F_{\rm drag}(\equiv\frac{1}{2}C_{\rm d}\rho A_{\rm b}U^2_{\rm term}),
\label{eq:balance}
\ee
where $V_{\rm b}$ and $A_{\rm b}$ are the volume and the horizontal cross-section area of the bubble, respectively; $\rho$ is the ICM density and $C_{\rm d}$  is the hydrodynamic drag coefficient. From Equation~(\ref{eq:balance}), the terminal velocity can be simply estimated to be
\be
U_{\rm term} = \sqrt{\frac{2gV_{\rm b}}{A_{\rm b}C_{\rm d}}}.
\label{eq:uterm}
\ee
Thus for a sphere,
\be
 U_{\rm term,\,sphere} = c_s\sqrt{\frac{4}{3\gamma C_{\rm d} } \frac{L}{H_{\rm P}}}\approx 1.3 c_s \sqrt{\frac{L}{H_{\rm P}}},
\ee
where for the last equality we used $C_{\rm d}\simeq0.5$, applicable for a sphere moving in incompressible fluid with the Reynolds number $\sim 10^3$ \citep{Roos1971}; $L$ is the diameter of the bubble. We can now estimate the Froude number
\be
  \Fr=\frac{U_{\rm term,\,sphere}}{NL}=\left( \frac{4\gamma}{3C_{\rm d}} \frac{H_{\rm S}}{L}\right)^{1/2}\approx 2.1 \sqrt{\frac{H_{\rm S}}{L}}.
 \label{eq:frs}
\ee
Thus, for plausible values of $L$ being a fraction of the scale height, the Froude number is larger than $1$, implying that internal-wave generation by a sphere is not going to be efficient. This is based on a number of simplifying assumptions, but the qualitative conclusion is that as long as simple hydrodynamic drag $F_{\rm hydro}$ dominates in $F_{\rm drag}\,(=F_{\rm hydro}+F_{\rm wave}+ {\rm...})$, the resulting $U_{\rm term}$ is high and the contribution of the wave drag $F_{\rm wave}$ is expected to be small. However, in the limit of $\Fr\gg1$, the wave drag $\propto\Fr^{-4}$ for a sphere in 3D \citep[see, e.g.][]{Gorodtsov2007}, implying that if one can reduce the terminal velocity, and consequently the Froude number, then the wave drag might become important.

Let us keep the assumption that $F_{\rm hydro}$ is the dominant term, but consider a flattened bubble, so that its vertical size is smaller than its horizontal size (see Fig.~\ref{fig:sketch}). For instance, consider a disk with diameter $L$ and thickness $h$, so that its frontal area $A_{\rm b}$ is $\pi L^2/4$ and the volume $V_{\rm b}=\pi L^2h/4$. Then the expression~(\ref{eq:uterm}) for the terminal velocity becomes
\be
  U_{\rm term,\,disk} = c_s\sqrt{\frac{2h}{\gamma C_{\rm d}H_{\rm P}}}\approx c_s\sqrt{\frac{h}{H_{\rm P}}},
\label{eq:utd}
\ee
where we use $C_{\rm d}\approx 1.2$ for a disk \citep{Roos1971}. Now only the vertical size $h$ of the bubble enters the expression. The corresponding Froude number is
\be
  \Fr=\frac{U_{\rm term,\,disk}}{NL}= \sqrt{\frac{2\gamma}{C_{\rm d}}} \sqrt{\frac{H_{\rm S}}{L}} \sqrt{\frac{h}{L}}\approx 1.7 \sqrt{\frac{H_{\rm S}}{L\epsilon_{\rm b}}},
\label{eq:frd}
\ee
where $\epsilon_{\rm b}\equiv L/h$ is the ratio of the horizontal and vertical dimensions of the bubble. Note that, when evaluating the Froude number, we use the horizontal size $L$ in the denominator, which is appropriate for $\epsilon_{\rm b}\gtrsim 1$. Comparing expressions (\ref{eq:frs}) and (\ref{eq:frd}), we see that in the former expression there is an extra factor $\epsilon_{\rm b}^{-1/2}$ that can make the Froude number smaller. Thus, for a given horizontal size $L$, the flattened bubble with $\epsilon_{\rm b}>1$ will have a smaller terminal velocity and a smaller Froude number. The condition that $\Fr\lesssim 1$ corresponds to $\epsilon_{\rm b}\gtrsim H_{\rm S}/L$. For a sufficiently large bubble with large $\epsilon_{\rm b}$, the excitation of internal waves might become important. Of course this would make the above estimates of the terminal velocity or the Froude number invalid, because they are based on the drag coefficient $C_{\rm d}$ corresponding to the unstratified case. It is expected that the drag will be higher due to the contribution of internal waves and other effects of the stratification and, therefore, the terminal velocity and the Froude number will be even lower accordingly. Thus, this exercise shows that flattened, buoyantly rising bubbles are potentially much more efficient at generating internal waves than spherical ones.
Here we emphasize again, that $U_{\rm term}$ itself is a function of $\Fr$, and, therefore, Equation~(\ref{eq:frd}) is expected to overestimate the true Froude number for a buoyantly rising body. It is therefore important to verify the qualitative statements made in this section with numerical simulations, presented in Section~\ref{sec:method}.

\begin{figure}
\centering
\includegraphics[width=0.45\textwidth]{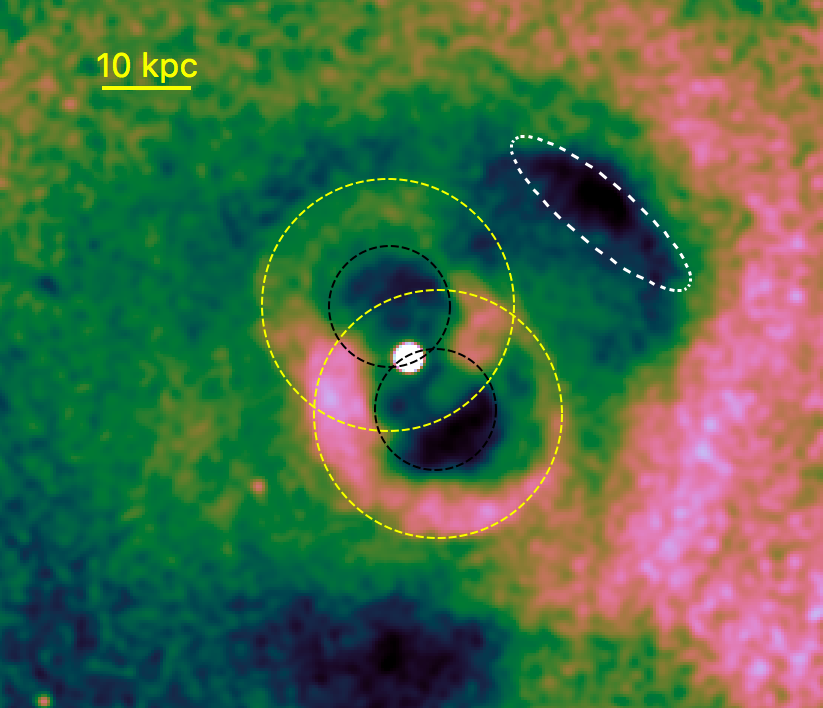}
\caption{\chandra 3.5-7.5 keV band image of the Perseus cluster. The bubbles appear as dark (X-ray dim) regions in this image. ``Active'' bubbles (radius $\sim7\kpc$) are marked with black dashed circles. They are surrounded by quasi-spherical weak shocks (radius $\sim14\kpc$), shown by yellow circles. The outer bubble, to the NW from the center, has the ``horizontal'' and ``vertical'' (radial) sizes $L\sim25\kpc$ and $h\sim7\kpc$, respectively. Thus, its aspect ratio is $\epsilon_{\rm b}=L/h\sim3.6$. We argue that for such bubbles the effects of stratification dominate in the total drag force.}
\label{fig:perseus}
\end{figure}

Is there any possible connection between the above discussion and the bubbles of relativistic plasma in clusters? A \chandra X-ray image of the Perseus cluster is shown in Figure~\ref{fig:perseus}. The radio bubbles are seen as dark areas (X-ray dim) in this image. The central two bubbles are more or less spherical. A weak shock that encompasses the bubbles indicates that they are still expanding \citep[see, e.g.,][]{Fabian2006,Zhuravleva2016,Tang2017}. However, older generations of bubbles, which are presumably rising buoyantly, are clearly aspherical. For example, shown with a white line is the ellipse with the ratio of major to minor axis $\sim 3.6$. While this ellipse is a poor match to the shape of the bubble observed at $\sim30\kpc$ to the North-West (NW) from the nucleus, it nevertheless provides a rough estimate of the degree to which bubbles might be flattened. The flattened bubble shape could be seen as a natural consequence of surface tension acting on the bubble surface. The ram pressure gradient of the flow squeezes the bubble along the direction of its motion, but the surface tension prevents the bubble surface from shredding. Putting aside the question of the nature of this surface tension, we note that substituting into Equation~(\ref{eq:frd}) the parameters of such a bubble, $H_{\rm S}\sim40\kpc$, $L\sim25\kpc$, $\epsilon_{\rm b}\sim3.6$, one gets $\Fr\lesssim1$. It is, therefore, plausible that the NW bubble can excite internal waves.

\section{The Model and Simulation Method} \label{sec:method}

In the previous section, we argued that the wave drag acting on flattened bubbles could be important, implying that a significant fraction of energy released during the buoyant rise would go into internal waves. In this section, we outline our program of testing this argument using numerical simulations. These simulations are not intended to reproduce closely the behavior of real bubbles in real clusters, but rather to demonstrate the ability of flattened bubbles to excite internal waves. To do this, we need the modeled bubbles to maintain their integrity as they move. In other words, we need a sufficiently strong surface tension to act on the bubble surface. While magnetic fields have been considered as a possible candidate to provide this surface tension \citep{Kaiser2005,Diehl2008,Bambic2018}, we have chosen to model the bubble as a rigid object and to use hydrodynamic simulations to investigate its dynamic behavior in the ICM (see Fig.~\ref{fig:sketch}). This choice dramatically simplifies the simulations and allows us to isolate the issue of internal-wave excitation from the much more difficult problem of determining the configuration of the magnetic field inside and outside the bubble. Given the qualitative character of this study, we perform 2D simulations, which corresponds to a flow past an elliptical cylinder in 3D.

Rigid bodies moving in a stratified atmosphere have been widely explored in industrial applications both
theoretically \citep[e.g.,][]{Warren1960,Voisin1994,Torres2000,Hanazaki2009} and experimentally \citep[e.g.,][]{Mowbray1967}. For the uniform vertical motion of a body, the Froude number is considered to be the key parameter, characterizing the generation of the internal waves \citep{Voisin2007}. This conclusion is broadly confirmed in our simulations.

\subsection{Simulation method and settings} \label{sec:method:simulation}
We address the problem in a two-dimensional (2D) Cartesian coordinate system $(x,\ y)$, and assume a fixed gravitational potential to model the cluster environment,
\be
  \Phi(x,\ y) = 2V_{\rm c}^{2}\ln(y+R_{\rm c}),
\label{eq:profile_pot}
\ee
where $V_{\rm c}$ and $R_{\rm c}$ are the scaling parameter for potential and the core radius in the unit of kpc, respectively. The plane-parallel ICM is initially assumed to be isothermal and in hydrostatic equilibrium. The corresponding gas density\footnote{Hereafter the variables with a zero in the subscript refer to the initial conditions.} is
\be
  \rho_{0}(x,\ y) = \rho_{\rm c}\exp\Big[-\frac{\Phi(x,\ y)}{{c_{T}}^2}\Big],
\label{eq:profile_rho}
\ee
where $\rho_{\rm c}$ and ${c_{T}}\equiv\sqrt{k_{\rm B}T_{\rm 0}/\mu m_{\rm p}}$ are the scale density and the isothermal sound speed of the gas, respectively; $T_{0}$, $k_{\rm B}$, $\mu\,(=0.6)$, and $m_{\rm p}$ are the initial gas temperature, Boltzmann constant, mean molecular weight per ion, and proton mass, respectively. The ICM is assumed to be an ideal monatomic gas with the adiabatic index $\gamma=5/3$. The dynamic viscosity of the ICM is set to $0.9\,{\rm g\,cm^{-1}\,s^{-1}}$, which is smaller than $5\%$ of the Spitzer value \citep{Braginskii1958,Spitzer1962}.

We model a 2D rigid bubble by an ellipse. Two parameters --- the horizontal length ($L$) and the ratio of the horizontal and vertical dimensions ($\epsilon_{\rm b}$) of the bubble --- define its size and shape. To obtain full-grown wakes and to investigate the effects of the bubble motion on the far-field regions, we set the computational domain to be $x\in[-150\kpc,\ 150\kpc]$ and $y\in[0,\ 400\kpc]$, and fix the initial bubble position at $(x_{\rm b0},\ y_{\rm b0})=(0,\ 40\kpc)$. Though in reality the interaction between the jets and the ICM could produce complex gas motions in the central region of the cluster while the bubble is formed, we assume that both the ICM and the bubble are initially static.

The force acting on the buoyant bubble is\footnote{Since we work in 2D geometry, the length of the infinitesimal bubble boundary element $d\ell$ is used in lieu of the surface element of a 3D body.}
\be
\mathbf{F}_{\rm bub} = \oint_{\rm bubble}{\left(P\mathbf{n}+\hat{\Pi}{\bf\cdot n}\right){\rm d}\ell},
\label{eq:force}
\ee
which contains the pressure and viscous components (i.e., the integral of the gas pressure $P$ and the shear stress $\hat{\Pi}$ over the bubble surface). Here $\bf n$ is the unit vector normal to the bubble surface. For simplicity, we only allow the bubble to move vertically, and do not consider the component of the force acting on the bubble in the horizontal direction (it should be zero for a symmetric flow). The gravitational force on the rigid bubble is also neglected, because the bubble mass density is assumed to be much smaller than that of the surrounding ICM.

\begin{figure}
\centering
\includegraphics[width=0.45\textwidth]{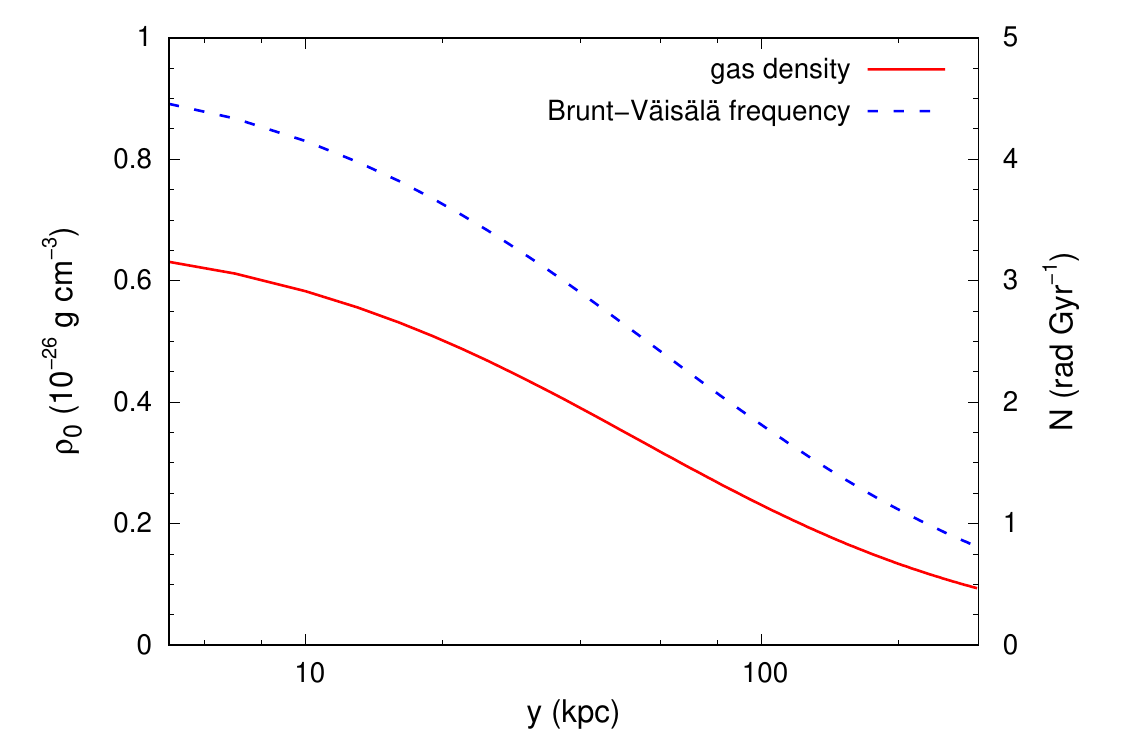}
\caption{Initial gas density profile (in red solid line, left axis, see Eq.~\ref{eq:profile_rho}) used in fiducial simulation runs, and the profile of the {\BV} frequency $N$ (in blue dashed line, right axis, see Eq.~\ref{eq:bv_frequency}). }
\label{fig:profiles}
\end{figure}

The main parameters of our simulations are summarized in Table~\ref{tab:ic_para}. The simulations can be split into three groups. In the first group, the bubble moves under the action of the buoyancy force in a stratified atmosphere. In the second group, the bubble moves with a constant vertical velocity $U$ in the same stratified atmosphere. In the third group, the bubble moves with a constant vertical velocity $U$ in a uniform atmosphere. The first group is the main content of this study, while the second and third groups are used to clarify the effect of the stratification and to verify the drag force dependence on a body's velocity. We use $V_{\rm c}=300\kms,\ R_{\rm c}=60\kpc,\ \rho_{\rm c}=6.77\times10^{-25}\,{\rm g\,cm^{-3}}$ and $k_{\rm B}T_{0}=1\keV$ for all our simulations except in the runs FE4L12highN and GE4L12V80highN. The initial gas density and the corresponding {\BV} frequency profiles are shown in Figure~\ref{fig:profiles}. The pressure and entropy scale heights at $y=100\kpc$ are $H_{\rm P}=140$ and $H_{\rm S}=210\kpc$, respectively. The runs FE4L12highN and GE4L12V80highN are performed to investigate the effect of a different gravitational field, where we select $V_{\rm c}=1000\kms,\ R_{\rm c}=60\kpc,\ \rho_{\rm c}=2.03\times10^{-21}\,{\rm g\,cm^{-3}}$ and $k_{\rm B}T_{0}=5\keV$, which provides a similar profile to that of Perseus cluster when $y>50\kpc$.

We use the OpenFOAM\footnote{Open Source Field Operation and Manipulation, www.openfoam.org} code to carry out our simulations. This code was designed for a wide range of applications of computational fluid dynamics, using the finite-volume method. Its flexibility to deal with non-orthogonal and moving meshes and the implementation of the subgrid turbulence models (e.g., Large Eddy Simulation technique, LES) make it highly suitable for our purposes. The effective spatial resolution of the simulations is $0.5\kpc$. We have confirmed the convergence of our simulations by increasing the resolution by a factor of two. The high-resolution run shows generally the same results as those from the default one. Further OpenFOAM settings adopted in this work are described in detail in Appendix~\ref{sec:appendix:simulation}.

\begin{table*}
\centering
\begin{minipage}{150mm}
\centering
\caption{Parameters of simulations}
\label{tab:ic_para}
\begin{tabular}{cccccccc}
  \hline\hline
 ID &
 $\epsilon_{\rm b}$\footnote{The ratio of the horizontal and vertical dimensions of the bubble $\epsilon_{\rm b}(\equiv L/h)$.} &
 $L\,(\kpc)$\footnote{The horizontal diameter of the bubble $L$.} &
 $U\,(\kms)$\footnote{The velocity of the bubble. For the bubbles that buoyantly float upwards (the `float' type), the velocity evolution is shown in Figure~\ref{fig:v_term}. } &
 Atmosphere\footnote{The environment where the bubble moves.}&
 Type\footnote{The motion type of the bubbles: buoyantly rising bubbles (the `float' type) and ones that move at fixed velocity (the `fix' type). }&\\
\hline
\multicolumn{6}{c}{Buoyancy driven bubbles in stratified atmosphere}\\
\hline
  FE1L6  & $1$ & 6  & - & stratified & float &  \\\hline
  FE1L12 & $1$ & 12 & - & stratified & float &  \\\hline
  FE4L6  & $4$ & 6  & - & stratified & float &  \\\hline
  FE4L12 & $4$ & 12 & - & stratified & float &  \\\hline
  FE4L24 & $4$ & 24 & - & stratified & float &  \\\hline
  FE4L48 & $4$ & 48 & - & stratified & float &  \\\hline
  FE4L12highN\footnote{The runs FE4L12highN and GE4L12V80highN used a different gravitational field from other simulations: it has a higher gas temperature and {\BV} frequency that are comparable to those of the Perseus cluster core. }
         & $4$ & 12 & - & stratified & float &  \\\hline
  FE9L12 & $9$ & 12 & - & stratified & float &  \\
\hline
\multicolumn{6}{c}{Bubbles moving with fixed velocity in stratified atmosphere}\\
\hline
  GE1L12V120 & $1$ & 12 & 120 & stratified & fix &  \\\hline
  GE4L12V15 & $4$ & 12 & 15 & stratified & fix &  \\\hline
  GE4L12V30 & $4$ & 12 & 30 & stratified & fix &  \\\hline
  GE4L12V60 & $4$ & 12 & 60 & stratified & fix &  \\\hline
  GE4L12V80highN & $4$ & 12 & 80 & stratified & fix &  \\\hline
  GE9L12V15 & $9$ & 12 & 15 & stratified & fix &  \\
\hline
\multicolumn{6}{c}{Bubbles moving with fixed velocity in uniform atmosphere}\\
\hline
  NE1L12V120 & $1$ & 12 & 120 & uniform & fix &  \\\hline
  NE4L12V30 & $4$ & 12 & 30 & uniform & fix &  \\\hline
  NE9L12V15 & $9$ & 12 & 15 & uniform & fix &  \\
\hline\hline
\vspace*{-2mm}
\end{tabular}
\end{minipage}
\end{table*}

\section{Results} \label{sec:results}

In this section, we present the main results of our simulations. Section~\ref{sec:results:buoyancy} describes a rigid bubble's buoyant rise through the ICM, and its impact on the ICM. Internal waves are generated downstream from the bubble. In Section~\ref{sec:results:comparison}, we make a comparison between bubbles moving in a stratified and an unstratified atmosphere. The stratification is found to play an important role in the drag coefficient. In Section~\ref{sec:results:dependence}, we study the dependence of the effective drag coefficient on the bubble's parameters and on the gravitational field. In Section~\ref{sec:results:analytic_model}, we compare the internal-wave patterns formed in our simulations with those predicted by linear theory.

\subsection{Buoyantly rising bubbles with different degrees of flattening} \label{sec:results:buoyancy}

\begin{figure*}
\centering
\includegraphics[width=0.9\textwidth]{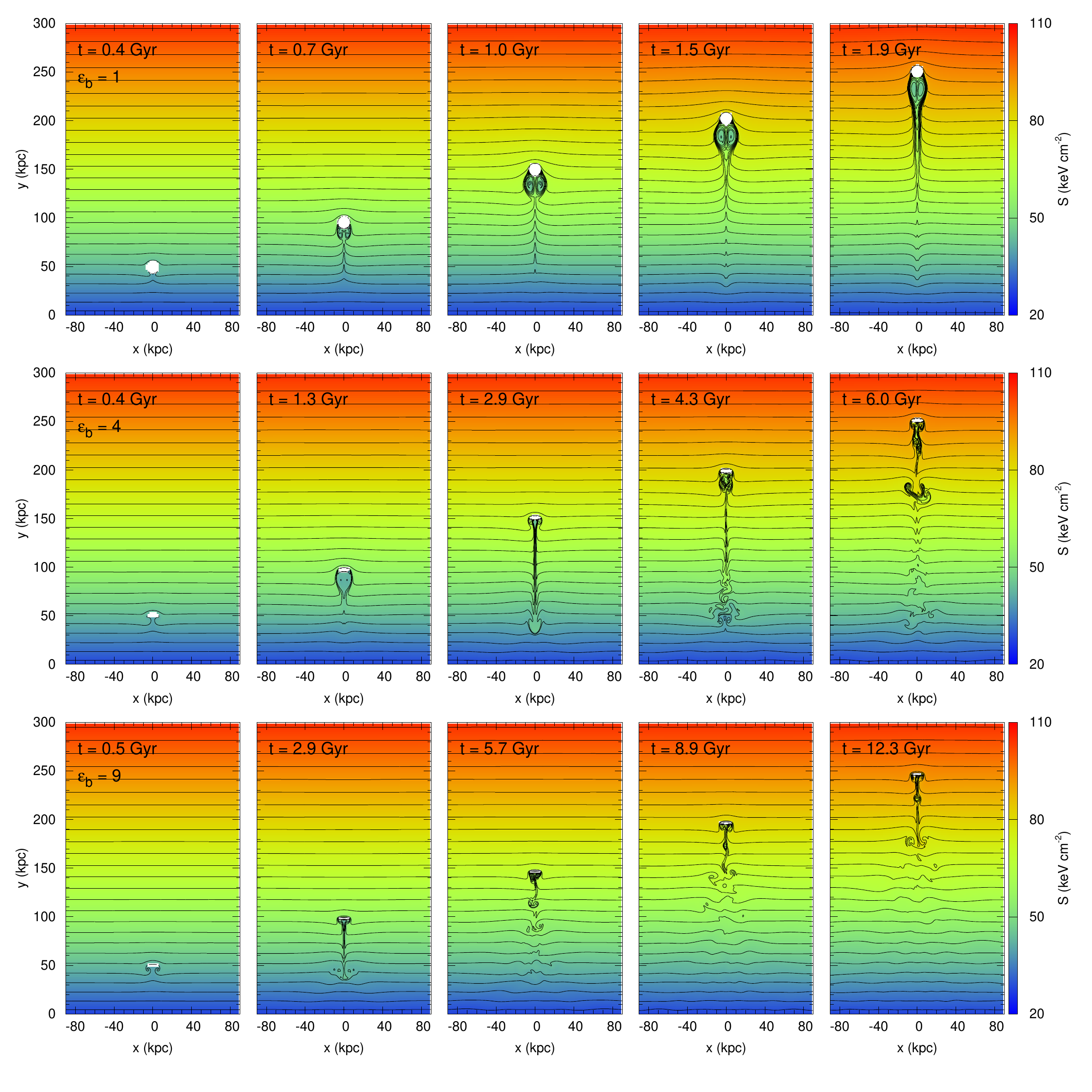}
\caption{Time evolution of the entropy maps for bubbles with different degrees of flattening
$\epsilon_{\rm b}=L/h$. Three  simulations are shown FE1L12 ($\epsilon_{\rm b}=1$, top row), FE4L12 ($\epsilon_{\rm b}=4$, middle row), and FE9L12 ($\epsilon_{\rm b}=9$, bottom row). The first one corresponds to a circular bubble, while in the other two the flattening is significant. The interval between each two successive contour levels in all the panels is $3\,{\rm keV\,cm^{-2}}$. The time elapsed since the beginning of the simulation is indicated in each panel.
This figure shows that the circular bubble (upper row) has the largest velocity and the perturbations induced in the ICM are confined to a narrow region trailing such a bubble. In contrast, flatter bubbles rise slower and cause perturbations over a larger region (see Section~\ref{sec:results:buoyancy}). }
\label{fig:entropy}
\end{figure*}

Figure~\ref{fig:entropy} shows the time sequences of the gas entropy map $S\,(\equiv T/\rho^{2/3})$ from the simulations FE1L12 (top row), FE4L12 (middle row), and FE9L12 (bottom row). In these simulations the bubbles, propelled by the buoyancy force, start to rise at $t=0$\footnote{A relaxation process for bubble acceleration is involved to reduce the effects of bubble's impulsive start (see Appendix~\ref{sec:appendix:simulation}).}. Figure~\ref{fig:entropy} illustrates the differences in the entropy perturbations induced by bubbles with different degrees of flattening, characterized by the parameter $\epsilon_{\rm b}$. For the circular-bubble case (i.e., $\epsilon_{\rm b}=1$, see the top row), a pair of vortices appears quickly on the downstream side of the bubble and survives until the end of the simulation. During the rise of the bubble, low-entropy gas is lifted up and forms a long and narrow channel trailing the moving bubble. As a result of the gravitational pull, the gas at the lower end of the channel decelerates after moving a distance $\sim10h$ (see the top panels at $t>1\Gyr$) and starts to fall back (it may further oscillate around equilibrium, although these oscillations are not visible in  Fig.~\ref{fig:entropy} due to the short duration of the run FE1L12). The key feature of these simulations is that the velocity of the bubble is high and all perturbations are confined to a narrow region in the wake of the bubble.

\begin{figure}
\centering
\includegraphics[width=0.45\textwidth]{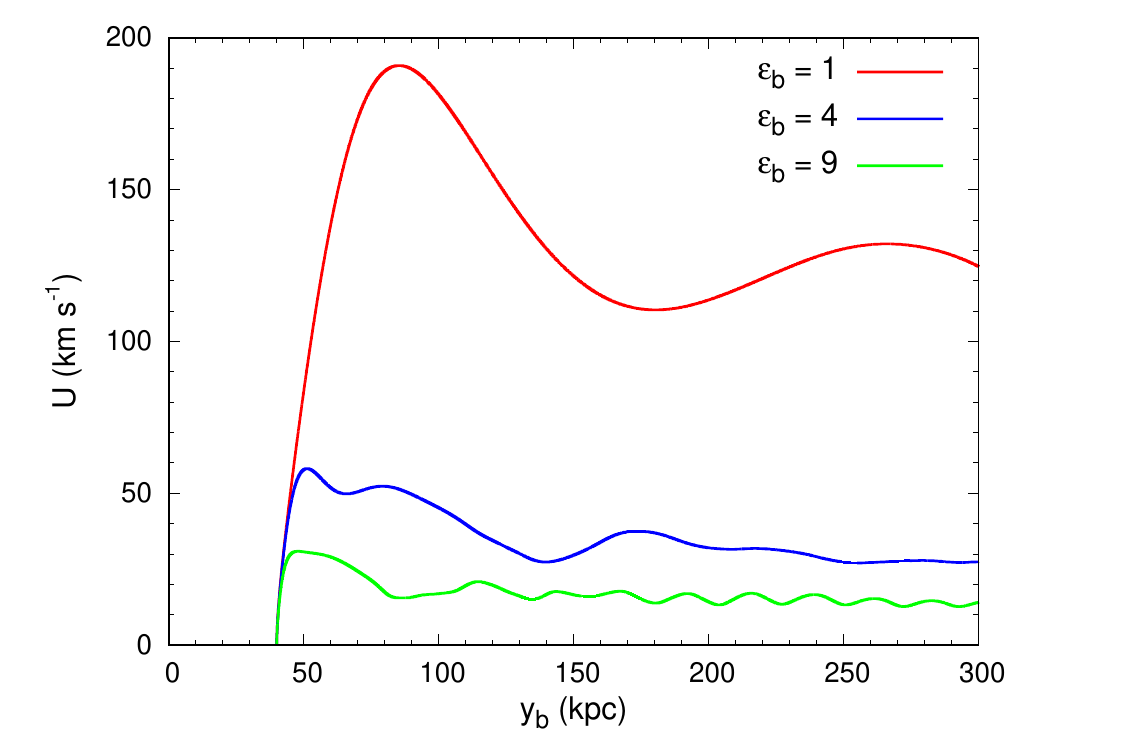}
\caption{Time evolution of the bubble-rise velocity in simulation FE1L12 (red line), FE4L12 (blue line), and FE9L12 (green line), respectively. The horizontal axis shows the vertical position of the bubble center $y_{\rm b}$, which is a monotonic function of time (similarly in Figs.~\ref{fig:ekin_frac} and \ref{fig:dc_cmp} below). The terminal velocity for the bubbles with $\epsilon_{\rm b}=1,\,4,\,9$ are $\sim120,\,30,\,15\kms$, respectively. As expected, flattened bubbles rise slower (see Section~\ref{sec:results:buoyancy}).}
\label{fig:v_term}
\end{figure}

Simulations of elliptical bubbles (see the middle and bottom rows in Fig.~\ref{fig:entropy}) show a very different picture. The bubbles rise much slower than in the run with a circular bubble (see Fig.~\ref{fig:v_term} for these rise velocities), while the vortices and the wake trailing the bubbles are much less prominent; the perturbations in the iso-entropy surfaces now extend in the horizontal direction over much larger distances. In the later evolutionary stages (i.e., after the bubble has risen $\sim50h$), the downstream vortices behind the elliptical bubbles become unstable, separate from the bubble surface, and shed downwards. We emphasize here that the vortex-shedding instability might be partly suppressed in our simulations, since the bubble is constrained to move vertically.

The evolution of the bubble velocities is shown in Figure~\ref{fig:v_term}. Independently of their shape, the bubbles experience a rapid acceleration in the early stage of the simulations. The bubbles first overshoot their respective terminal velocities, but then decelerate by the pressure drag\footnote{The physical viscous drag is much smaller than the pressure drag in our simulations, though it is still included in the calculations.} and settle approximately at the terminal velocity.  Figure~\ref{fig:v_term} shows clearly that flattened bubbles rise with smaller terminal velocities. We note that the terminal velocity is not exactly constant in the simulations, which mildly depends on the shape of the potential well. In our simulations, bubbles generally require $\sim2\Gyr$ to achieve the terminal velocity. This is related to the value of the relaxation factor $\beta$ in the numerical scheme used for calculation of the bubble acceleration (see Appendix~\ref{sec:appendix:simulation}). The terminal velocities of the bubbles with different degrees of flattening are $120\kms\ (\epsilon_{\rm b}=1)$, $30\kms\ (\epsilon_{\rm b}=4)$ and $15\kms\ (\epsilon_{\rm b}=9)$.

The slower rise velocities of the strongly flattened bubbles in the simulations do not necessarily imply the effect of stratification. Indeed, in these simulations the horizontal size of the bubble was kept constant, while  $\epsilon_{\rm b}$ increased from 1 to 9, leading to a decrease of the bubble volume ($\propto 1/\epsilon_{\rm b}$) and, consequently, of the buoyancy force. Since the frontal area of the bubble does not change, the normal hydrodynamic drag (at a given velocity) varies only as long as the drag coefficient varies. In 2D, the difference in the drag coefficient between the ``sphere'' and the ``disk'' is less than a factor of 2. Therefore, even in the absence of stratification effects, one can expect that when the driving force (the buoyancy force in our case) decreases by a factor of $\sim 9$, the terminal velocity of a towed body will also decrease. We will elucidate the contribution of the stratification to the drag in the next section. Here we only note that the presence of internal waves can already be seen in Figure~\ref{fig:entropy}. Generation of internal waves by bubbles generally proceeds via two channels: (1) the bubble can directly excite the internal waves (body-driven generation); (2) the bubble first excites eddies within the wake (and uplifts the gas in the wake), which then generate the internal waves (wake-driven generation). At low $\Fr$, the former channel becomes more important \citep[][]{Brandt2015}. The simulations FE4L12 and FE9L12 were run for a long enough time ($>8\Gyr$) to see the long-term signatures of interaction between the bubble and the ambient ICM. One can clearly see the signs of internal waves in the middle and bottom panels of Figure~\ref{fig:entropy}, viz., the wave-like pattern revealed in the iso-entropy surfaces or the collapse of the eddies far from the bubble (see also Figs.~\ref{fig:ekin_map}, \ref{fig:ekin_evolution} and \ref{fig:ekin_vcmp} below, for a clearer view of the internal waves).

\begin{figure}
\centering
\includegraphics[width=0.45\textwidth]{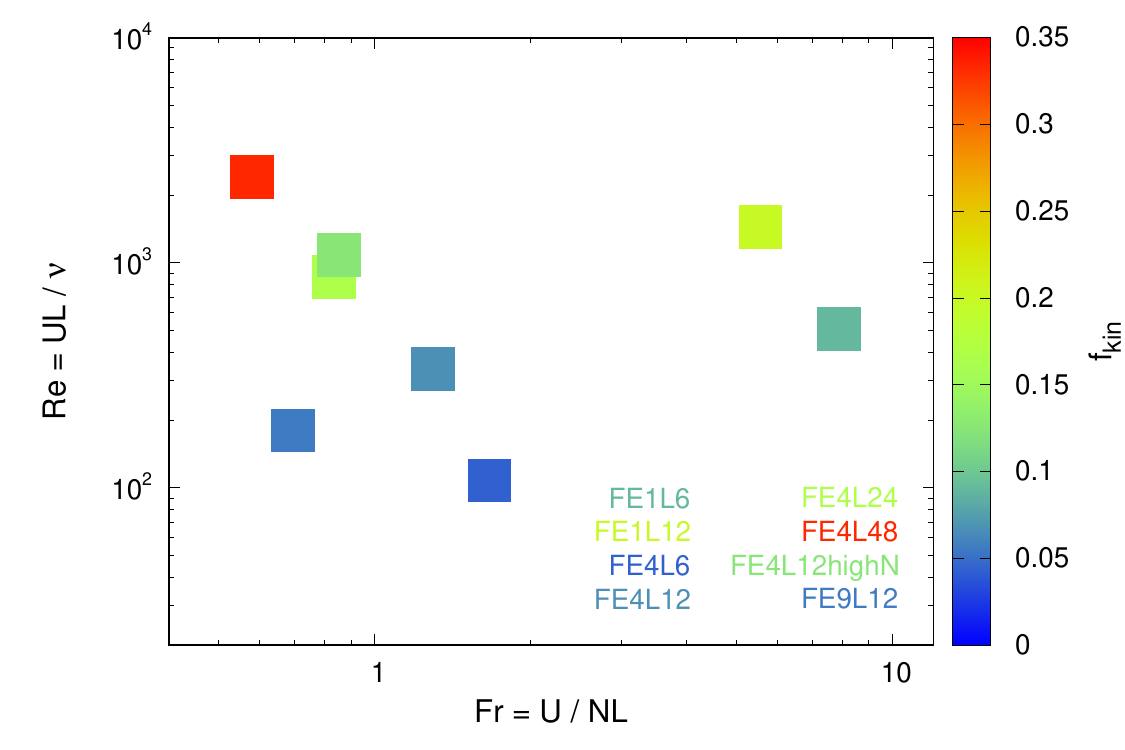}
\caption{The kinetic fraction (defined by Eq.~\ref{eq:f_kin}) as a function of the Froude number and Reynolds number from different sets of simulations when bubble arrives at $y=250\kpc$. The kinematic viscosity is set to be $\nu=3\times10^{26}{\rm\,cm^2\,s^{-1}}$ in all runs; the {\BV} frequency is taken at the height $y=150\kpc$, viz., $N=8.8\,{\rm rad\,Gyr^{-1}}$ (for FE4L12highN) and $N=1.8\,{\rm rad\,Gyr^{-1}}$ (for all other runs). This figure shows
the parameter region $(\rm\Fr,\ Re)$ our simulations have covered,  and suggests that for smaller bubbles the  energy dissipation is significant (see Section~\ref{sec:results:buoyancy}). }
\label{fig:fr_re}
\end{figure}

The fraction of energy that goes to internal waves via the eddy excitation may depend on the viscosity and diffusivity of the fluid (both physical and numerical), since the excitation of waves by eddies is competing with the viscous dissipation and mixing of the gas within each eddy. In other words, one can expect the Reynolds number to have significant effect on the solution \citep[see, e.g., fig.~2 in][]{Torres2000}. To assess the level of dissipation in our simulations, we have calculated the ratio of the kinetic energy present in the simulation box to the total energy deposited by the bubble
\be
f_{\rm kin} = \frac{\int\int\rho(x,\ y)u^2(x,\ y)/2{\,\rm d}x{\rm d}y}{\int F_{\rm buoyancy}{\,\rm d}y},
\label{eq:f_kin}
\ee
where $F_{\rm buoyancy}=\rho_{0}gV_{\rm b}$ is the buoyancy force, $u(x,\ y)$ is the gas velocity field. Note that, the total energy associated with gas motions can be higher due to the contribution of the potential energy. For instance, for internal waves the contribution of potential energy doubles the energy fraction in the waves. The value of $f_{\rm kin}$ is shown (color-coded) in  Figure~\ref{fig:fr_re} as a function of the Froude and the Reynolds  ${\rm Re}=UL/\nu$ numbers,
when the bubble arrives at $y=250\kpc$. For the definition of the Reynolds number we used the value of physical kinematic viscosity set in the code $\nu \approx 3\times10^{26}{\rm\,cm^2\,s^{-1}}$. As we show in the next section, our simulations lead to the appearance of a ``Christmas tree'' pattern formed by the internal waves in the wake of the bubble. According to the chart compiled by \cite{Torres2000}, such patterns are characteristic for flows with $\Fr\sim 1$ and $\rm Re\sim a\ few\ 100$, in agreement with Figure~\ref{fig:fr_re}.  This figure shows that $f_{\rm kin}$ ranges $\sim5\%-35\%$ for the simulation parameter sets we have covered, being mainly dependent on the Reynolds number. For a bubble with a given shape, the simple scaling discussed in Section~\ref{sec:preliminary} predicts that $U\propto \sqrt{L}$, i.e. the Reynolds number increases with $L$ as $L^{3/2}$, while the Froude number decreases as $L^{-1/2}$. Indeed, Figure~\ref{fig:fr_re} suggests that for smaller bubbles more energy is dissipated relatively quickly, while for larger bubbles most of the released energy survives for longer time (especially given that kinetic energy accounts only for a part of the energy).

\citet{Torres2000} simulated a sphere vertically moving in a uniformly stratified fluid. They found a caudal fluid column formed by the collapse of the vortices behind the body, with maximum velocity in the column up to ten times that of the sphere \citep[see also][]{Hanazaki2009}. This rear jet-like structure moving in the opposite direction to the sphere leads to a significant increase of the pressure drag acting on the body. In our simulations, we do see a long narrow channel trailing the circular bubble (together with a pair of vortices), which however has a velocity comparable to that of the bubble. It might not correspond to the jet-like structure found in \citet{Torres2000}, whose formation seems to depend on the thin boundary layers near the body surface, which may not be well resolved in our simulations.

\subsection{Comparison of bubbles moving in stratified and uniform atmospheres} \label{sec:results:comparison}

We have seen that buoyantly rising bubbles, in particular flattened ones, cause perturbations in the ICM outside the wake region. To confirm that these perturbations are indeed internal waves and to clarify the role of these waves in spreading the energy over the volume, we make a direct comparison between simulations of bubbles moving with a fixed velocity in stratified and uniform atmospheres. For each value of $\epsilon_{\rm b}$, we use the terminal velocity obtained in Section~\ref{sec:results:buoyancy} for buoyantly rising bubbles with the corresponding degrees of flattening (see Table~\ref{tab:ic_para} for the simulation parameters)\footnote{To reduce the effect of an impulsive start, we make the bubbles start with zero velocity and accelerate them artificially to the preset velocity $U_{\rm term}$ over $\sim0.5\Gyr$. }.

\subsubsection{Energy flow in the far field}

\begin{figure*}
\centering
\includegraphics[width=0.7\textwidth]{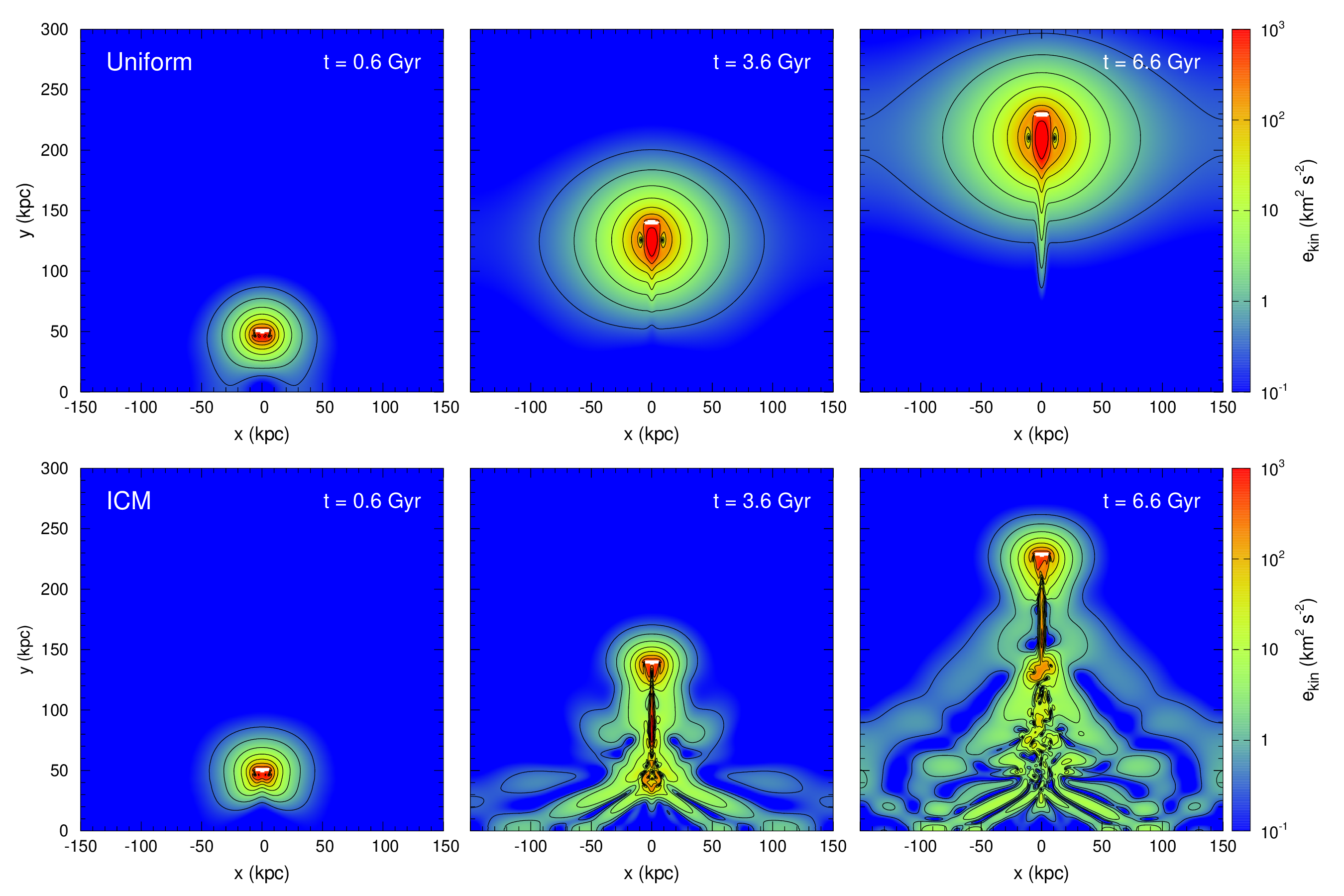}
\caption{Comparison of the specific kinetic energy of the gas in simulations with bubbles moving in unstratified (run NE4L12V30, top panels) and stratified (run GE4L12V30, bottom panels) atmospheres. In the stratified case, internal waves are excited, revealed by a characteristic ``Christmas tree'' pattern (this pattern is also known as `flared skirt' or `herring bone'). Clearly, a significant amount of the undissipated energy is transported to the far field by the internal waves (see Section~\ref{sec:results:comparison}).}
\label{fig:ekin_map}
\end{figure*}

The specific kinetic energy of the gas in stratified and unstratified runs, viz., $u(x,\ y)^{2}/2$, for an elliptical bubble with $\epsilon_{\rm b}=4$ is shown in Figure~\ref{fig:ekin_map}. The time-averaged potential and kinetic energies of linear waves are equal. Therefore, the specific kinetic energy is a good tracer of the energy flow associated with linear waves in the system. As seen in the top panels, in an unstratified atmosphere, most of the kinetic energy is confined to the vortices and the surrounding area. The bubble perturbs its local field and a narrow channel along its path. In the far-field region (say $|x|>2L$), the gas is barely perturbed.
In contrast to the uniform-atmosphere run, in the stratified atmosphere, the bubble makes a significant impact on the far-field region (downstream of the bubble, see the bottom panels of Fig.~\ref{fig:ekin_map}). The excited internal waves reveal themselves as a ``Christmas tree'' pattern.

\begin{figure}
\centering
\includegraphics[width=0.45\textwidth]{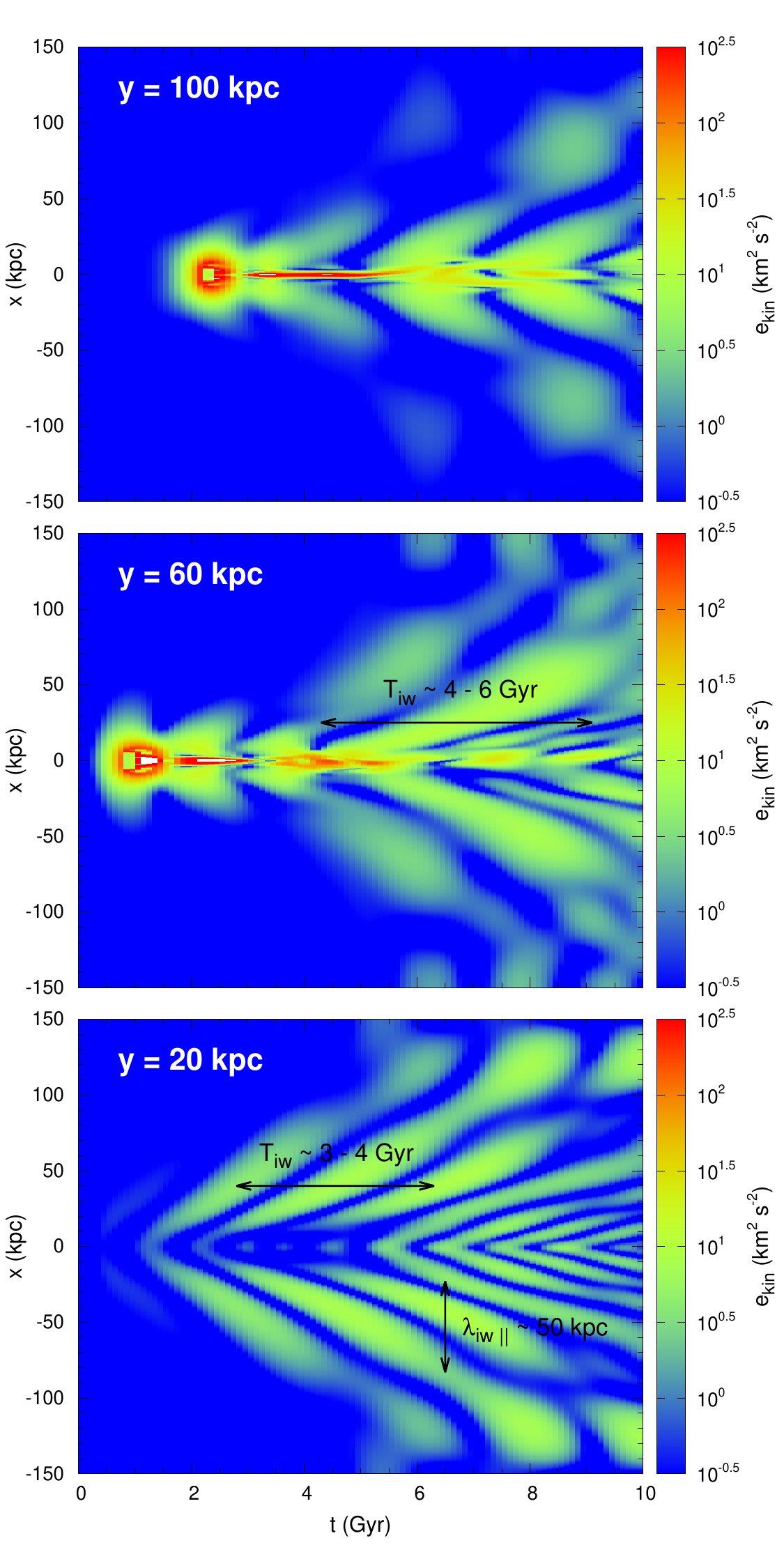}
\caption{Time evolution of 1D slices of the specific kinetic energy of the gas at $y=20\kpc$ (bottom panel), $60\kpc$ (middle panel), and $100\kpc$ (top panel) for the run GE4L12V30. The periodic pattern reveals internal waves propagating outwards (up and down in this plot). The period of internal waves is approximately $T_{\rm iw}\sim3-4\Gyr$ at $y=20\kpc$ and $T_{\rm iw}\sim4-6\Gyr$ at both $y=60$ and $100\kpc$ (marked in the panels, see Section~\ref{sec:results:comparison}).}
\label{fig:ekin_evolution}
\end{figure}

Figure~\ref{fig:ekin_evolution} shows the time evolution of 1D horizontal slices of the specific kinetic energy at $y=20,\ 60,\ 100\kpc$ for the run GE4L12V30. The periodic pattern of crests and troughs shows that the excited internal waves propagate in the horizontal direction away from the wake of the bubble. We can directly read the period $T_{\rm iw}$ and the wavelength (the reciprocal of the horizontal wavenumber) $\lambda_{iw\,||}$ of the internal waves from these time-space images (the resulting values are marked in the panels). For example, the pattern seen in the bottom panel (at $y=20\kpc$) has the period $T_{\rm iw}\sim3-4\Gyr$. The corresponding frequency $\omega_{\rm iw}(\equiv2\pi/T_{\rm iw})$ is a factor of $\sim 0.5$ smaller than the {\BV} frequency at this height. The wavelength of the internal wave at $y=20\kpc$ is $\lambda_{iw\,||}\sim50\kpc$. At $y=60$ and $100\kpc$, the period of waves becomes longer (i.e., $T_{\rm iw}\sim4-6\Gyr$).

\begin{figure}
\centering
\includegraphics[width=0.45\textwidth]{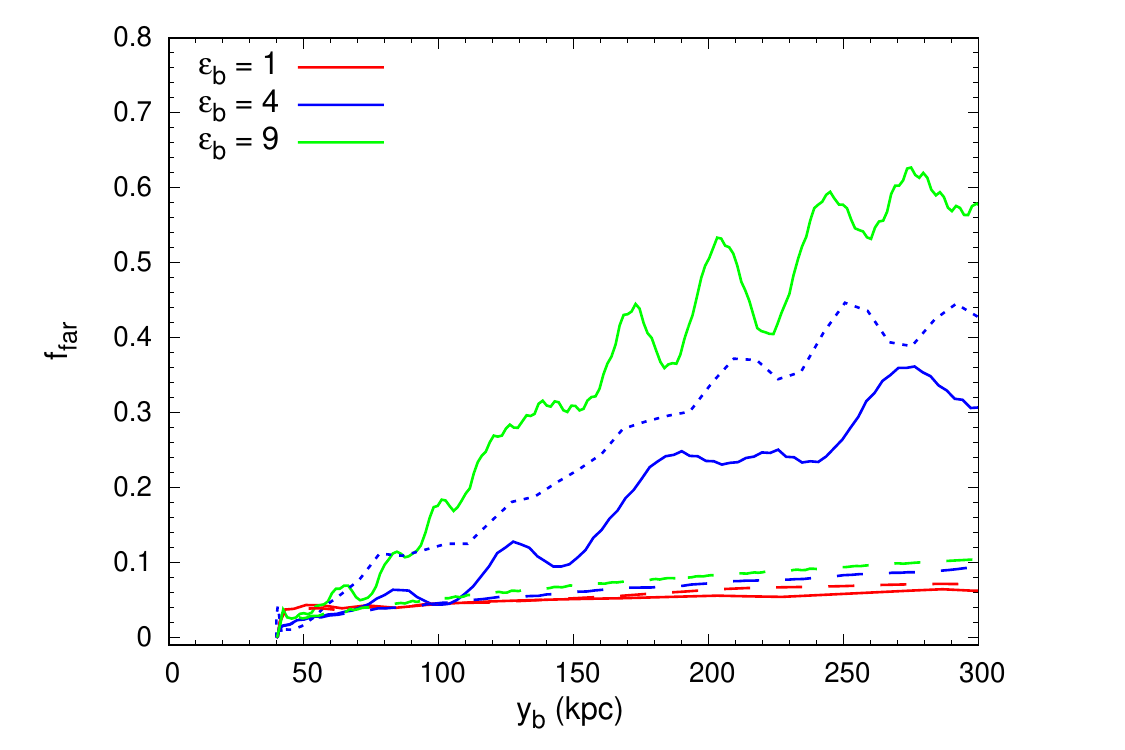}
\caption{Time evolution of the far-field kinetic energy fraction $f_{\rm far}$ (defined by Eq.~\ref{eq:f_far}) for bubbles with different degrees of flattening, moving in stratified (solid lines, runs GE1L12V120, GE4L12V30, GE9L12V15, respectively) and unstratified (dashed lines, runs NE1L12V120, NE4L12V30, NE9L12V15, respectively) atmospheres. The result from run GE4L12V80highN is also plotted for a comparison (blue dotted line). This figure shows that internal waves are efficient in transporting energy from the bubble into the far field (see Section~\ref{sec:results:comparison}).}
\label{fig:ekin_frac}
\end{figure}

Using the kinetic-energy maps, we can estimate the kinetic energy fraction in the far field, viz.,
\be
f_{\rm far} = \frac{\int_{|x|>2L}\int\rho(x,\ y)u^2(x,\ y){\rm d}x{\rm d}y}{\int\int\rho(x,\ y)u^2(x,\ y){\rm d}x{\rm d}y},
\label{eq:f_far}
\ee
to quantify the efficiency of the energy flow away from the bubble wake. A homogeneous distribution of energy corresponds to $f_{\rm far}=(x_{\rm max}-2L)/x_{\rm max}\approx 0.84$, where $x_{\rm max}=150\kpc$ is the horizontal half size of the simulated box. The value $\sim0.8$ would imply that the kinetic energy is efficiently transported in the horizontal direction. The time evolution of $f_{\rm far}$ for the bubbles with different $\epsilon_{\rm b}$ is shown in Figure~\ref{fig:ekin_frac}. For the unstratified cases (see the dashed lines), the fractions are lower than $10\%$ throughout the simulation, and show little dependence on the bubble shape. Thus, in the absence of stratification, the energy does not spread away from the wake in the horizontal direction.

In contract, the internal waves formed in a stratified atmosphere provide an efficient way for transporting their energy in horizontal direction (see the solid lines in Fig.~\ref{fig:ekin_frac}). The evolution of the energy fraction in the stratified runs shows significant dependence on the bubble flattening (viz., Froude number). For example, the kinetic-energy fractions for the bubbles with $\epsilon_{\rm b}=1,\ 4,\ 9$ are approximately $6\%,\ 30\%,\ 60\%$, respectively, when the bubble arrives at $y=300\kpc$.

Combining with the results presented in Figure~\ref{fig:fr_re}, we conclude that, according to our simulations,
$\sim5\%-30\%$ of the energy released by bubble is propagated into the far field. Given that the quantitative answer depends on the Reynolds number and, also, that our simulations are done in 2D, we plan to explore a wider parameter range $\rm (Fr,\ Re)$ and the effects of 3D geometry in future studies.

\subsubsection{Increase of effective drag coefficient in stratified medium}

\begin{figure}
\centering
\includegraphics[width=0.45\textwidth]{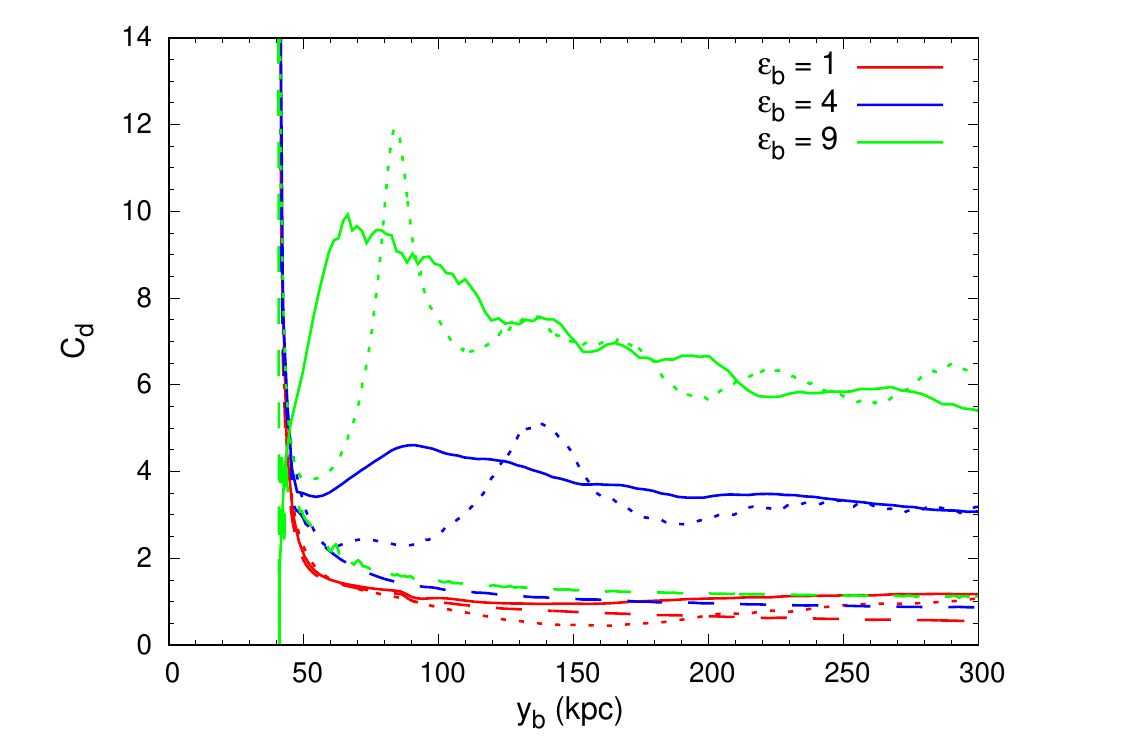}
\caption{The drag coefficients $C_{\rm d} $ for bubbles with different $\epsilon_{\rm b}$ moving at constant velocity in stratified (solid lines, runs GE1L12V120, GE4L12V30, GE9L12V15, respectively) and unstratified (dashed lines, runs NE1L12V120, NE4L12V30, NE9L12V15, respectively) atmospheres. The drag coefficient for buoyantly rising bubbles is also plotted for a comparison (dotted lines, runs FE1L12, FE4L12, FE9L12, respectively). This figure shows the dramatic effect of the stratification on the drag coefficient for strongly flattened bubbles (see Section~\ref{sec:results:comparison}).  }
\label{fig:dc_cmp}
\end{figure}

We now proceed with the evaluation of the effective drag coefficient for bubbles moving in a stratified medium. Of course, the contribution to the drag coefficient could come from the excitation of the internal waves, advected low-entropy gas, viscous drag and turbulence generated in the wake of the bubble. The total force $F_{\rm bub}$ acting on the bubble (see Eq.~\ref{eq:force}) that rises at terminal velocity is zero. It can be decomposed into two components: the drag force (including the pressure and viscous drags) and the buoyancy force:
\be
F_{\rm bub}=F_{\rm drag}+F_{\rm buoyancy}= 0.
\ee
We evaluate the stratification-dependent drag coefficient as
\be
C_{\rm d}(\Fr) = \frac{F_{\rm drag}}{\rho_{0}(x_{\rm b},\,y_{\rm b}) A_{\rm b}U^2/2},
\label{eq:dc}
\ee
where $U$ is the velocity of the bubble and $(x_{\rm b},\,y_{\rm b})$ are the coordinates of the bubble's center, and substitute $F_{\rm drag}$ with the buoyancy force $F_{\rm buoyancy}$. We can separate the drag coefficient into two contributions (see \citealt{Voisin2007}): one without the stratification $C_{\rm d}(\Fr=\infty)$ and the remaining part that contains the effects of stratification,
\be
C_{\rm d}(\Fr) = C_{\rm d}(\Fr=\infty) + \Delta C_{\rm d}(\Fr).
\label{eq:dc_f}
\ee
In Figure~\ref{fig:dc_cmp}, we compare the evolution of the drag coefficient for bubbles moving with the same velocities in stratified and unstratified atmospheres. For the unstratified cases, the drag coefficient $C_{\rm d}(\Fr=\infty)$ ranges between $\sim0.6$ and $\sim1.2$ (shown with dashed lines), which mildly depends on the bubble shape. The time-averaged drag coefficient of an infinitely long circular cylinder, as reported in the literature (obtained from numerical simulations or measured in experiments, see, e.g., \citealp{Cantwell1983,Breuer1998}), is $\sim1.2$, which is larger than our estimate. This might be due to two reasons: (1) we do not treat the bubble boundary in such details as the studies focused on measuring the drag coefficient accurately, which require much higher resolution near the bubble surface; (2) the flow passing a blunt body (not a Stokes flow) is usually unsteady: large-scale vortex shedding in our simulations takes place at later times $t>50h/U$, so, the system is still in the transition phase when we evaluate the drag coefficient. Despite these issues, our simulations suffice for the purposes of this study, given that the deviations from the values reported in the literature are much smaller than the contribution to the drag caused by the stratification. The solid lines in Figure~\ref{fig:dc_cmp} show the drag coefficient when the bubble moves at constant velocity (approximately equal to the terminal velocity of the bubble) in a stratified atmosphere. One can clearly see the prominent effect of stratification on the drag coefficient, which has a significant dependence on the bubble shape, viz., $\Delta C_{\rm d}(\Fr) \sim0.5,\,2.5,\,5.0$ for the bubbles with the shape parameter $\epsilon_{\rm b}=1,\,4,\,9$, respectively. For comparison, we also show the drag coefficients evaluated for the buoyantly rising bubbles (shown with the dotted lines in Fig.~\ref{fig:dc_cmp}). Overall, $\Delta C_{\rm d}(\Fr)$ for buoyantly moving bubbles follows the corresponding values for the cases with constant velocity equal to the mean terminal velocity, but with obvious oscillations, caused by unsteady motion of the buoyantly rising bubbles.

\subsection{Dependence of effective drag coefficient on bubble parameters and gravitational field} \label{sec:results:dependence}

\begin{figure}
\centering
\includegraphics[width=0.45\textwidth]{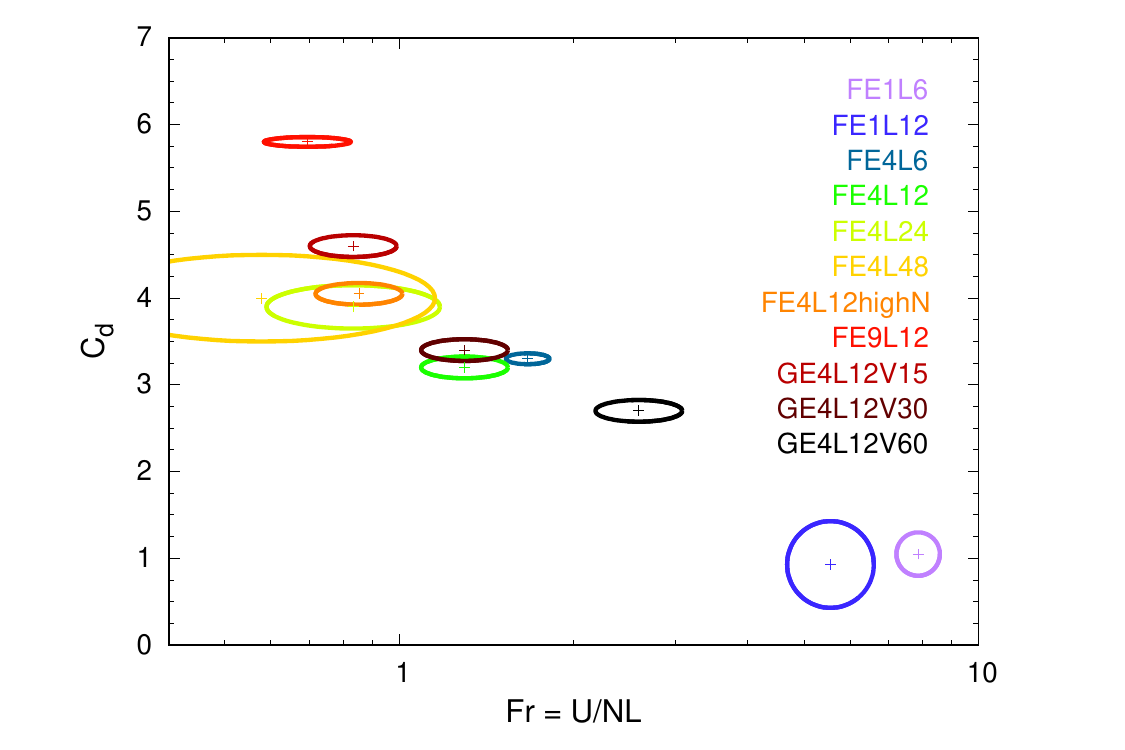}
\caption{Comparison of the drag coefficient from different sets of the simulations (see Table~\ref{tab:ic_para}). The {\BV} frequency is taken at the height $y=150\kpc$, viz., $N=8.8\,{\rm rad\,Gyr^{-1}}$ (for FE4L12highN) and $N=1.8\,{\rm rad\,Gyr^{-1}}$ (for all other runs). The ellipses illustrate the shape and the relative size of bubbles used in the simulations. This figure shows how the drag coefficient depends on the bubble characteristics (i.e., shape $\epsilon_{\rm b}$, size $L$, and velocity $U$) and on the level of stratification $H_{\rm S}$ and $g$ (see Section~\ref{sec:results:dependence}). }
\label{fig:dc_dep}
\end{figure}

Figure~\ref{fig:dc_dep} shows the dependence of the effective drag coefficient on the bubble parameters (size, shape, and velocity) and on the gravitational field for a bubble moving in a stratified medium. The purpose of this plot is to verify if the estimated values of the drag coefficient in our simulations can be extrapolated to more realistic cases without doing additional simulations (see discussion of the Perseus and M87/Virgo clusters in Section~\ref{sec:discussions}). Two trends are clear from Figure~\ref{fig:dc_dep}. First, fixing the bubble size and shape but varying the velocity, shows that the drag coefficient goes down with the bubble velocity. In other words, $\Delta C_{\rm d}(\Fr)$ in Equation~(\ref{eq:dc_f}) is a decreasing function of the Froude number. Secondly, when the bubble rises buoyantly, the effective drag coefficient is a strong function of the bubble shape, but shows only weak dependence on the bubble size and on the degree of stratification, i.e., $H_{\rm S}$ or $g$. Therefore, we can assume that the effective drag coefficient of a buoyantly rising bubble is only a function of the bubble shape $\epsilon_{\rm b}$. This conclusion, of course, is based on a limited set of simulations, but suffices for the qualitative consideration adopted here. As a direct inference, the terminal velocity is proportional to the square root of the bubble size, i.e., $U_{\rm term}\propto\sqrt{h}$. Therefore, for two bubbles of the same shape, the larger one would rise at higher terminal velocity.

\subsection{Qualitative comparison of simulated pattern of internal waves with linear theory} \label{sec:results:analytic_model}

\begin{figure*}
\centering
\includegraphics[width=0.7\textwidth]{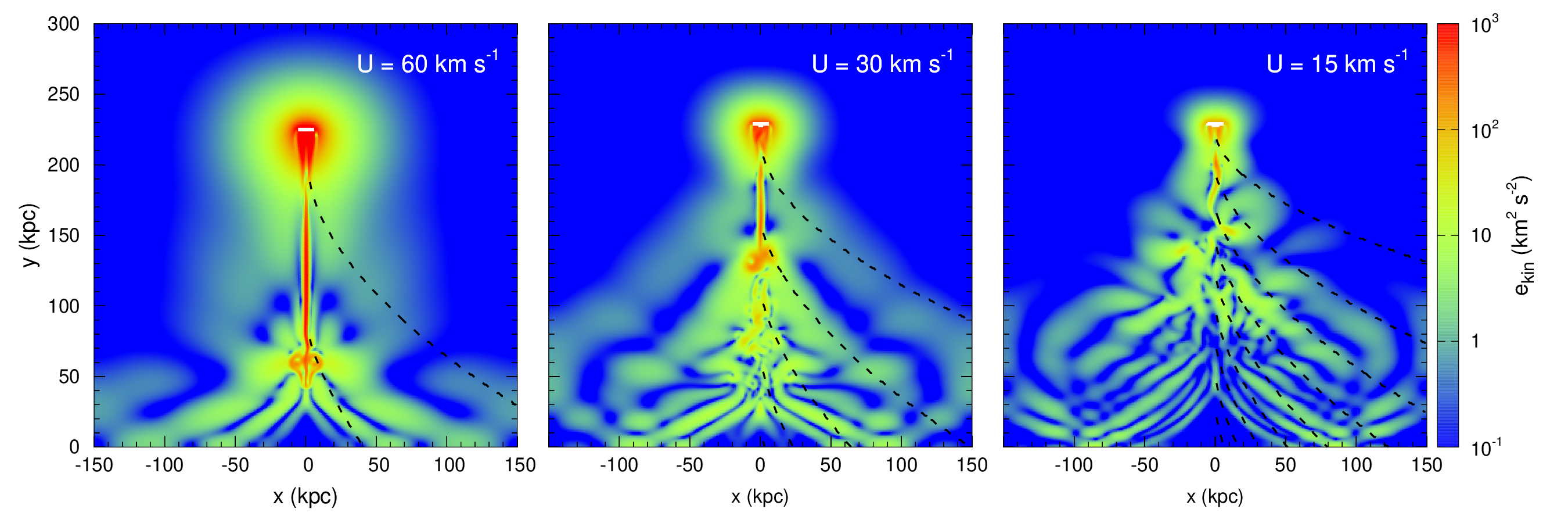}
\caption{Surfaces of constant phase in the linear approximation superposed onto the map of the specific kinetic energy in our simulations. The legends are similar to those for Figure~\ref{fig:ekin_map}, but for a comparison of the bubble ($\epsilon=4$) moving in the stratified atmosphere at different velocities, i.e., $U=60,\,30,\,15\kms$. The snapshots are selected from the runs GE4L12V60, GE4L12V30, GE4L12V15, in such a way as to capture the bubble when they have arrived at the same height. The dashed lines represent the surfaces of constant phase of the internal wave predicted by the analytical model. This figure shows that the analytical model of \citet{Voisin1991,Voisin1994} qualitatively reproduces the pattern of the internal waves observed in  simulations (see Section~\ref{sec:results:analytic_model}).}
\label{fig:ekin_vcmp}
\end{figure*}

\citet{Voisin1991,Voisin1994} used linear theory to predict analytically the pattern of internal waves generated by a towed 3D body in a uniformly stratified incompressible fluid. In this section, we make a qualitative comparison between his predictions and our simulations. Figure~\ref{fig:ekin_vcmp} shows the specific kinetic energy of the gas for simulations with bubbles moving at different velocities ($U=60,\,30,\,15\kms$) but with fixed size ($L=12\kpc$) and shape ($\epsilon_{\rm b}=4$). The surfaces of constant phase of a body-driven internal wave estimated from the linear theory for incompressible fluid (see more details in Appendix~\ref{sec:appendix:analytic_model}) are over-plotted as the dashed lines. For calculating these surfaces, we used a dipole source to mimic the moving bubble, rather than considering its specific shape. This approximation is valid when the wavelength of the internal wave is much longer than the body size. As is seen from Figure~\ref{fig:ekin_vcmp}, the linear model agrees qualitatively with the pattern of internal waves found in our simulations. However, the pattern is more bent and concentrated in the simulations near the bottom boundary, due to the reflective boundary condition and the variant {\BV} frequency $N$ used in the simulation (see Fig.~\ref{fig:profiles}, but constant $N=2$ used in the analytic model). In the linear theory, the wavelength of excited internal waves is approximately proportional to the bubble velocity $\lambda_{\rm iw}\propto U/N$. Thus, the distribution of troughs and crests is expected to be more concentrated when the bubble velocity is smaller, consistent with the simulation results. In addition, the linear theory provides an explanation of the weak dependence of the gas velocity on the bubble velocity in the far field. In the linear theory, the gas velocity is $u(x,\ y)\propto Nm_{\rm d}/Ur_{\rm h}^2$, where $m_{\rm d}$ is the dipole moment proportional to the volume and velocity of the source (see Eq.~\ref{eq:dipole}), $r_{\rm h}$ is the horizontal distance from the source. The $u(x,\ y)$ is therefore not a function of the bubble velocity.

Obviously, the non-linear features seen in the simulations (e.g., the wake) could not be captured by the linear model. Figure~\ref{fig:ekin_vcmp} shows that stronger turbulence is generated in the wake of the bubble when it rises with higher velocity (i.e., larger $\Fr$ and $\rm Re$). In this case, a larger fraction of the kinetic energy is associated with the wake and the uplifted gas. %Therefore, at larger $\Fr$, the wake-driven internal waves are expected to contribute more to the energy transport.

\section{Discussion} \label{sec:discussions}

By considering a rigid bubble buoyantly rising in a stratified medium, we have found that flattened bubbles can efficiently excite internal waves, which possibly make an important contribution to the total drag that the ICM exerts on the bubble. These waves tap the potential energy of the rising bubble and transport this energy horizontally away from the bubble's wake. Given that in clusters internal waves are trapped in the radial direction by the decreasing {\BV} frequency, the waves must eventually dissipate and heat the ICM.

This conclusion depends critically on the ability of the bubbles to maintain their integrity. For instance, in simulations of \citet{Reynolds2015}, explosive energy release in localized volumes led to the appearance of a shock-heated bubbles of the gas, which were rapidly shredded by hydrodynamic instabilities and mixed with the surrounding ICM. No efficient excitation of the internal waves was seen (or could be foreseen) and heating took place locally via mixing of the hot gas in the bubbles with the ambient ICM (cf., \citealt{Hillel2014}).

Guided by the results of the simulations discussed in the previous section, we consider the implications of the buoyantly rising (flattened) bubbles on the heating of the ICM in the context of the radio-mode AGN feedback.

\subsection{Constraints on bubble terminal velocity in Perseus cluster and M87} \label{sec:discussions:vterm}

\begin{figure}
\centering
\includegraphics[width=0.45\textwidth]{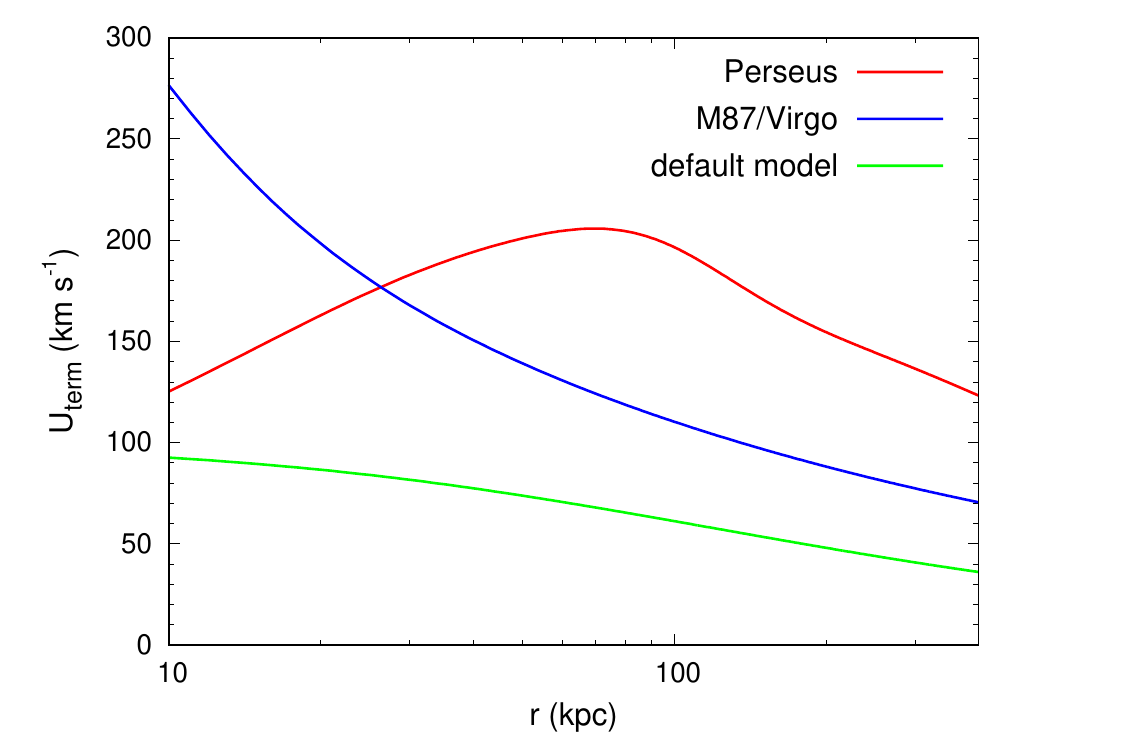}
\caption{Estimated terminal velocity as a function of the cluster radius for Perseus, M87/Virgo and our default model used in the simulations (from Eq.~\ref{eq:uterm}), assuming $\epsilon_{\rm b}=4$ and $V_{\rm b}/A_{\rm b}=5\kpc$, for which $C_{\rm d}=3$.  In the Perseus, the expected terminal velocity of the NW bubble is $\sim100-200\kms$  (see Section~\ref{sec:discussions:vterm}). }
\label{fig:vterm_cluster}
\end{figure}

In our simulations, we have found that for bubbles with different degrees of flattening (i.e.,
$\epsilon_{\rm b}=1-9$), the effective drag coefficients are in the range $C_{\rm d}\simeq1-7$ (see Sections~\ref{sec:results:comparison} and \ref{sec:results:dependence}). The weak dependence of the drag coefficient on the bubble's size and the degree of stratification of the ICM allows us to estimate the bubble terminal velocity in galaxy clusters by combining Equation~(\ref{eq:uterm}) and the observed radial profiles of density and pressure. The estimated bubble terminal velocities as functions of radius for the Perseus cluster and M87 are shown in Figure~\ref{fig:vterm_cluster}. For these calculations, we used the radial profiles measured via the X-ray observations (see appendix~F in \citealt{Tang2017} and references therein). The shape and the height of the bubble are fixed to $\epsilon_{\rm b}=4$ and $V_{\rm b}/A_{\rm b}=5\kpc$ (motivated by the parameters of the NW bubble in the Perseus cluster shown in Fig.~\ref{fig:perseus}), for which $C_{\rm d}=3$. For the Perseus cluster, the estimated terminal velocity is in the range $\sim150-200\kms$ within $r<400\kpc$. This velocity corresponds to $\sim 0.2c_{\rm s}$, a factor of $2-3$ slower than that assumed in \citet{Churazov2000,Churazov2001}, whose estimates were made for a rising spherical bubble.

The morphology of the wake formed behind the rising bubble could provide an independent way of constraining the rise velocity of the buoyant bubbles. In the simulations shown in Figure~\ref{fig:entropy}, the vertical size $l$ of the vortex behind the bubble is $\simeq75,\,30,\,15\kpc$ for the bubbles with $\epsilon_{\rm b}=1,\,4,\,9$, respectively. These simulations suggest that one can roughly estimate $l$ from the condition that $U/Nl\sim1$. Therefore, the rise velocity of the buoyant bubbles in galaxy clusters could be estimated as $Nl$ if we know the size of the vortex. This could be inferred, for instance, from the horseshoe-shaped H$\alpha$ filaments found in the Perseus cluster \citep{Conselice2001,Fabian2003,Hatch2006} trailing the NW bubble, shown in Figure~\ref{fig:perseus} (see also fig.~3 in \citealt{Fabian2003}). The size of the region occupied by these filaments is $l\simeq20-30\kpc$, and therefore the rise velocity $U\simeq200-300\kms$. This estimate agrees with the results shown in Figure~\ref{fig:vterm_cluster}. Moreover, the \citet{Hitomi2017} reported the velocity dispersion $\sigma_{\rm v}=203^{+16}_{-15}\kms$ around the region of the NW bubble, while the corresponding bulk velocity was relatively small: $v_{\rm bulk}=60^{+20}_{-21}\kms$. All these results broadly agree with the scenario that the NW bubble rises close to the plane of the sky with velocity $\sim200\kms$.

On a more speculative note, it is possible to associate the absorption feature trailing the NW bubble in Perseus cluster with the formation of a fluid rear-jets in the bubble's wake, caused by stratification. In simulations and some laboratory experiments \citep[see, e.g.,][]{Torres2000}, the velocities in the direction opposite to the body motion were up to ten times the body's velocity. While formation of such jets may depend on the level of stratification, the shape of the bubble and the properties of the ICM, it is tempting to consider the possibility that the so-called High Velocity System \citep{Minkowski1957} might originate from such mechanism.

\subsection{Velocity diagnostic with future high-energy-resolution missions}

Future missions like \athena \citep[e.g.,][]{Nandra2013} and \lynx \citep[e.g.,][]{Gaskin2016}, which offer a combination of high angular and energy resolutions, will dramatically change our ability to probe the velocity field in the cool-core regions. Although a clean direct detection of internal waves generated by individual bubbles in the far field might be challenging due to projection effects and interference of patterns from several bubbles, the velocities in the near field might offer a cleaner test of the effects of stratification. In particular, the simplest test might be done by measuring the magnitude and the spatial extent of the bubble-induced velocities ahead of the bubble. If one neglects stratification, the velocity field ahead of a flattened body might resemble a potential flow affecting a region with size of the order of the transverse size of the body. Strong stratification works towards restoring the isopycnal (or isoentropic) surfaces, thus pushing these surfaces closer to the body. The result is a flow that has a smaller radial extent compared to the unstratified case, as clearly seen from the comparison of the top and bottom panels in Figure~\ref{fig:ekin_map}. At the same time, the horizontal velocity increases to compensate for the reduced cross-section of the flow. Therefore, for a bubble rising approximately in the plane of the sky, mapping the velocity field ahead of the bubble could allow one to determine both the level of the stratification and the bubble's rise velocity. Another interesting diagnostic could arise from studying the flow behind the bubble, which is sensitive to the combination of the degree of stratification and to effective viscosity \citep[see, e.g.,][]{Torres2000}. We defer quantitative predictions for these effects to a future publication.

\subsection{Energy flow in cool cores} \label{sec:discussions:number_bub}

If bubbles of relativistic plasma are capable to maintain their integrity as they rise buoyantly, then the following simple scenario describes the overall energy flow in cool cores of relaxed clusters. The vertical transport of energy is provided by rising bubbles, which transfer energy to internal waves. These internal waves spread the energy over a larger volume of the ICM, principally in the tangential, rather than radial, direction. Since internal waves are trapped in the core of the cluster \citep{Balbus1990}, they become nonlinear and dissipate \citep[e.g.,][]{Zhuravleva2014}. Given that bubbles lose their energy after crossing a few pressure scale heights, the efficiency of coupling their energy to the ICM is very high, which is the characteristic feature of the rising-bubble scenario \citep[][]{Churazov2001,Churazov2002,Begelman2001}. The reduced terminal velocity of flattened bubbles implies that more bubbles are expected in the cool cores of a cluster at any given moment. However, in the steady-state scenario, the total dissipation rate does not depend on the rise velocity of the bubbles, but is equal to the rate at which the AGN deposits energy into the bubbles.

\section{Conclusions} \label{sec:conclusions}

We have shown that buoyant bubbles of relativistic plasma, ubiquitously found in the cores of galaxy clusters, can be efficient emitters of internal waves. A sufficient condition for this is a substantial flattening of the bubbles, so that their vertical (radial) size is a factor of a few times smaller than the azimuthal one. Their terminal velocity scales as the square root of their sizes, but the effective drag coefficient is larger than that for a spherical bubble. For such flattened bubbles, the terminal velocities are small enough so that the Froude number $\Fr\lesssim 1$. The stratification effects make the dominant contribution to the total drag force balancing the buoyancy force and clear signs of internal waves are seen in these simulations. The internal waves propagate horizontally and downwards from the rising bubble, spreading the energy over large volumes of the ICM, where it eventually dissipates. As in the majority of other buoyancy-controlled scenarios, the coupling efficiency of the AGN mechanical power with the ICM is large and the energy does not leak outside cluster cores. When scaled to the conditions of the Perseus cluster and M87, the expected terminal velocity is $\sim100-200\kms$ near the cluster cores (assuming the bubble vertical size of $5\kpc$). For the Perseus cluster, the velocities are in broad agreement with direct measurements by the Hitomi satellite.

Finally, we emphasize the qualitative nature of the model considered here. We used 2D simulations and modeled the bubbles as rigid bodies, rather than solving a self-consistent problem, which would require a much more sophisticated description of the stratified atmosphere and of the bubble itself. Nevertheless, we believe that the concept of flattened bubbles correctly captures the most important features of the proposed model, namely increased drag coefficient and, as a consequence,  more efficient generation of the internal waves.  We defer a study of the 3D geometry and probing broader parameter space in the $\rm(Fr,\ Re)$ plane for future work.

\section*{Acknowledgements}
EC acknowledges support by grant No.~14-22-00271 from the Russian Scientific Foundation. The work of AAS was supported in part by grants from UK STFC and EPSRC.

\appendix
\section{The OpenFOAM settings}
\label{sec:appendix:simulation}

Here we summarize the detailed OpenFOAM settings (including the solver, the mesh, and the subgrid turbulence model) used in our simulations.
\begin{itemize}
  \item We performed our simulations by using the built-in rhoPimpleDyMFoam solver, which was developed for dealing with compressible flows. We modified this solver by adding a static gravitational field (see Eq.~\ref{eq:profile_pot}).
  \item The snappyHexMesh tool was applied to generate the mesh for the computational domain. The bubble was modeled as a moving wall with a no-slip boundary condition. Reflecting boundary condition was adopted at the edge of the domain in both of the $x$ and $y$ directions. At the bottom and top boundaries, the `fixedFluxPressure' condition is used for the pressure to keep the stability of the atmosphere.
  \item The LES approach was employed in our simulations. This can efficiently reduces the computational cost by subgrid modeling of the small-scale turbulence instead of directly simulating it. We select the one-equation eddy-viscosity model (kEqn) to handle the subgrid turbulence. The turbulent energy and the modified turbulent viscosity were uniformly set to be $k=10^{11}\,{\rm cm^{2}\,s^{-2}}$ and $\tilde{\nu}=3\times10^{26}\,{\rm cm^{2}\,s^{-1}}$ in the initial conditions. We further tested different subgrid turbulence models (e.g., the k-omega-SST-DES model) and model parameters, and found that they had little impact on our results.
\end{itemize}

To reduce the effects of an impulsive start of the bubble and improve computational stability, we included a relaxation process for bubble acceleration. This limits the amount of acceleration changing from one time step to the next, i.e., $a_i = (1-\beta)a_{i-1} + \beta\hat{a}_i$, where $a_{i-1}$ and $a_{i}$ are the bubble acceleration at the step $i-1$ and $i$; $\hat{a}_i$ is the expected acceleration obtained from Equation~(\ref{eq:force}) at the step $i$; and $\beta\,(=10^{-4})$ is the relaxation factor. The relaxation affects bubble motion until the bubble velocity approaches its terminal velocity; it also filters out fast variations in the bubble velocity. Since we mainly focus on excitation of internal waves by bubbles moving at the terminal velocity and on time-averaged quantities (e.g., the effective drag coefficient), this artificial relaxation does not affect our conclusions.

\section{Analytical model}
\label{sec:appendix:analytic_model}

Here we summarize the linear theory \citep{Voisin1991,Voisin1994} applied in Section~\ref{sec:results:analytic_model} to predict analytically the pattern of internal waves generated by bubbles\footnote{This theory is developed for 3D cases.}. Assuming an incompressible and uniformly stratified (i.e., of constant {\BV} frequency $N$) fluid and including the Boussinesq approximation, \citet{Voisin1994} found a static solution of the Green's function for the far-field velocity distribution $u_{\rm mono}$ in the case of the uniform vertical motion of a monopole source $m_{0}$ (in the rest frame of the source, see his eq.~6.9):
\be
u_{\rm mono}\propto\frac{Nm_{0}}{Ur_{h}}\cos\Big[{\frac{N}{|U|}}f(r_{h},\ r_{z})\Big],
\label{eq:monopole}
\ee
where $f(r_{h},\ r_{z})$ is a function of the horizontal $(r_{h})$ and vertical $(r_{z})$ distance from the source. In this study, we consider a dipole source to model the finite volume of the moving bubble:
\be
u_{\rm dipole}=\frac{m_{\rm d}}{m_{0}}\frac{{\rm d} u_{\rm mono}}{{\rm d}r_{z}},
\label{eq:dipole}
\ee
where $m_{\rm d}\propto a^{3}U$ is the dipole moment, $a$ is the effective radius of the sphere modeled by the dipole. This approximation is valid in the far field where the wavelength of the internal waves is much longer than the bubble size $L$ (see Fig.~\ref{fig:ekin_evolution}). The surfaces of the constant phase of $u_{\rm dipole}$ are shown in Figure~\ref{fig:ekin_vcmp}, where $N$ is fixed to $2$.

\bsp	% typesetting comment
\label{lastpage}
\end{document}